\documentstyle[11pt]{article}\textheight 230mm\textwidth 150mm
            \newcommand{\be}{\begin{eqnarray}}
            \newcommand{\ee}{\end{eqnarray}}
           \newcommand{\eel}[1]{\label{#1}\end{eqnarray}}
\newcommand{\e}[1]{\label{e:#1}\end{eqnarray}}
     \newcommand{\eg}{{\em e.g.\ }}
            \newcommand{\ie}{{\em i.e.\ }}
            \newcommand{\ga}{{\gamma}}
 
            \newcommand{\la}{{\lambda}}
       \newcommand{\om}{{\omega}}
            
            \newcommand{\del}{{\delta}}
   \newcommand{\dq}{\dot{q}}
   \newcommand{\dt}{\dot{t}}

           \newcommand{\ra}{{\rightarrow}}

            \newcommand{\pet}{{\cal P}}
\newcommand{\bapet}{\bar{\cal P}}
\newcommand{\tbapet}{\tilde{\bar{\cal P}}}

 \newcommand{\tpet}{\tilde{\cal P}}
 \newcommand{\tca}{\tilde{{\cal C}}}

 \newcommand{\hpet}{\hat{\cal P}}
 \newcommand{\hca}{\hat{{\cal C}}}
 \newcommand{\hbapet}{\hat{\bar{\cal P}}}
 \newcommand{\hbaca}{\hat{\bar{\cal C}}}

\newcommand{\ca}{{\cal C}}
\newcommand{\baca}{\bar{\cal C}}
 \newcommand{\tbaca}{\tilde{{\baca}}}
            \newcommand{\tpsi}{\tilde{\psi}}
            \newcommand{\beq}{\begin{quote}}
            \newcommand{\eq}{\end{quote}}
            \newcommand{\Om}{\Omega}
            \newcommand{\al}{\alpha}
            \newcommand{\ben}{\begin{enumerate}}
            \newcommand{\een}{\end{enumerate}}
            \newcommand{\bit}{\begin{itemize}}
            \newcommand{\ei}{\end{itemize}}
    	\newcommand{\nn}{\nonumber}
            \newcommand{\r}[1]{(\ref{e:#1})}
            \newcommand{\edfl}[1]{\Label{#1}\end{df}}

\newcommand{\vb}{{\cal h}}
\newcommand{\hb}{{\cal i}}

\newcommand{\ve}{{\varepsilon}}

\newcommand{\dagg}{^{\dag}}
\newcommand{\sign}{\mbox{sign}}
\newcommand{\bett}{{\bf 1}}
\newcommand{\dif}{{\partial}}
\newcommand{\half}{\frac{1}{2}}

\begin{document}
\begin{titlepage}
\noindent
June 22, 2000\\

\vspace*{5 mm}
\vspace*{20mm}
\begin{center}{\LARGE\bf
Quantum BRST properties of \\reparametrization invariant theories.}\end{center}
\vspace*{3 mm}
\begin{center}
\vspace*{3 mm}

\begin{center}Robert
Marnelius\footnote{E-mail: tferm@fy.chalmers.se} and
 Niclas Sandstr\"om\footnote{E-mail: tfens@fy.chalmers.se}
 \\ \vspace*{7 mm} {\sl
Institute of Theoretical Physics\\ Chalmers University of Technology\\
G\"{o}teborg University\\
S-412 96  G\"{o}teborg, Sweden}\end{center}
\vspace*{25 mm}
\begin{abstract}
Any regular quantum mechanical system may be cast into an abelian gauge theory
by simply reformulating it  as a reparametrization invariant theory.
We present a detailed study of the BRST
quantization of such reparametrization invariant theories  within a precise
operator
version of BRST.  The treatment elucidates several intricate aspects
of the BRST quantization of reparametrization invariant theories like the
appearance
of physical time. We propose general rules for how physical wave functions and
physical propagators are to be projected from the BRST singlets and
propagators in
the ghost extended BRST theory. These projections are performed by boundary
conditions which are precisely specified by the operator BRST. We  demonstrate
explicitly the validity of these rules for the considered class of models. The
corresponding path integrals are worked out explicitly and compared with the
conventional BFV path integral formulation.
\end{abstract}\end{center}\end{titlepage}

\setcounter{page}{1}
\setcounter{equation}{0}
\section{Introduction.}
The main purpose of the present paper is to obtain a precise understanding
of the
quantum properties of reparametrization invariant theories as they appear
in a BRST
quantization. We want to know precisely how physical objects may be
extracted and in
particular how physical time appears from the formalism.  We restrict
ourselves to
the framework of ordinary quantum mechanics with finite number of degrees
of freedom
since  we are  interested in the generic case without topological obstructions.

That any regular classical Lagrangian mechanical system may be cast into an
equivalent
reparametrization invariant theory has been known for long \cite{Lan,HT}. The
procedure is as follows:
Consider a regular Lagrangian theory with the Lagrangian $L(t)\equiv
L(\dq(t),\dq(t);t)$. ($L(t)$ is regular if
$\det{\dif^2L(t)\over\dif\dq^i\dif\dq^j}\neq0$.) Consider then the action
and replace
time
$t$ by an arbitrary parameter $\tau$:
\be
&&\int dt L(t)=\int d\tau\: \dt L(t(\tau)),\quad \dt\equiv{dt\over d\tau}>0.
\e{101}
The new Lagrangian $L'(\tau)\equiv\dt L(t(\tau))$ does then obviously
describe the same theory. However, $L'(\tau)$ contains time $t(\tau)$ as a new
degree of freedom and is singular. Due to the invariance under
reparametrizations
$\tau\ra\tau'(\tau)$  we have the constraint
\be
&&\pi+H(t)\equiv {\dif L'\over\dif\dt}+H(t)=0,
\e{102}
where $H(t)$ is the Hamiltonian of the original theory described by $L(t)$.
$H(t)$
 depends explicitly on time when $L(t)$ does. In this way we have turned the
original regular theory into an abelian gauge theory with the constraint variable
$\pi+H(t)$ as gauge generator.  In  appendix A  we derive all relevant
observables,
\ie all relevant gauge invariant objects of this classical gauge
theory. The physical time is shown to be just the
gauge fixed time.

In a Dirac quantization of the above reparametrization invariant theory the
Schr\"odinger equation appears as  a pure constraint equation. The equivalence
with the original theory is then obvious. (For a recent application see
\cite{Tkach}.)
The gauge invariant variables in  appendix A may be turned into operators
by means
of which formal solutions and propagators of the Schr\"odinger equation may be
constructed. (A corresponding
detailed treatment of the relativistic particle with and without spins were
given in
\cite{RMBN}.)

In the present paper we  perform a BRST quantization of
the above reparametrization invariant theory. We  call it
cohomological quantum mechanics to distinguish it from the quantization of the
original Lagrangian. It has been partly treated in \eg
\cite{HT,GVU,Time}. However, here we  perform a very detailed analysis of this
 BRST quantization based on a precise operator formulation on inner product
spaces, a formulation which also was used in \cite{Time}. We shall
explicitly demonstrate the relation to ordinary quantum mechanics which is
much more
involved than in the Dirac quantization above. In the process we also find the
necessary ingredients of the BRST theory for this correspondence to exist.   We
consider
 general gauge fixings and  general state representations. {In
particular, we discuss in detail how time appears from the reparametrization
invariant version}. (The problems with physical time in reparametrization
invariant
theories, particular gravity, is discussed in \cite{Ish}.) In addition we
propose
general projection formulas for
 physical wave functions and
 propagators, which for cohomological quantum mechanics may be directly
compared with the corresponding objects in the original quantum theory.

The paper is organized as follows: In section 2 we present the precise operator
version of  the BRST quantization which we then use for  cohomological quantum
mechanics in section 3. In section 4 we consider possible wave function
representations of a set of  BRST singlets. We propose general formulas for
physical wave functions and propagators which are shown to yield the
expected results
for this set. In section 5 we consider another set and
discuss the differences. In section 6  we give and calculate path integral
expressions for propagators, compare them with the conventional BFV
formulation and
show how they may be projected. Finally we conclude the paper in section 7. In
appendix A we give gauge invariant extensions for the classical version of the
reparametrization invariant  theory. In appendix B we give some properties
of our
ghost states and in appendix C we apply our projection formulas to the
relativistic
particle.

\setcounter{equation}{0}
\section{Our BRST approach.}
There are several roads to  a satisfactory quantization of general gauge
theories (see \cite{HT} for a review). The general approach is called
 BRST quantization which is  cohomological in
nature.  There are two versions. One is  Hamiltonian
 and is called BFV-BRST quantization \cite{BFV}, and the other  is
Lagrangian and is
called field-antifield BV-quantization \cite{BV}. Both were originally
designed for
path integral quantizations. In the BFV-BRST approach one identifies the
classical
constraints and introduces generalized Faddeev-Popov ghosts to construct a
nilpotent
odd BRST charge in a Poisson sense. BFV has given a general prescription
for this
charge. They have also a general prescription for the BRST invariant Hamiltonian
which has to be inserted into a phase space version of the path integral. The
BV-approach on the other hand has a completely different form but should
yield an equivalent result at the end. Both these methods yield an
effective theory.
However, to handle these effective theories one  has often to do some modifications
due to the necessity to work on inner product spaces. In quantum field
theory one
typically  has to consider a Euclidean version of the effective theory.

In the present paper we shall make use of a precise operator formulation of BRST
quantization on inner product spaces based on the BFV scheme. It allows for a
detailed treatment of the quantum theory and provides for algorithms for the
physical states.  Any operator BRST quantization requires
\cite{KO}
\be
&&Q|ph\hb=0,\quad Q^2=0,
\e{1}
where $|ph\hb$ is any physical state, and where $Q$ is the odd, hermitian
BRST charge
operator which lives in a ghost extended framework. In this framework the
operators
and states may be decomposed into eigenoperators and eigenstates to the
ghost number
operator
$N$. In particular we have $[N, Q]=Q$, which means that $Q$ has ghost
number one. To
construct a nilpotent
$Q$ one may  use  the general BFV-prescription for the pseudoclassical $Q$ and
then quantize. For finite degrees of freedom one always arrives in this way
 at a nilpotent operator
$Q$ by means of an expansion in
$\hbar$ \cite{BF}. However, for infinite degrees of freedom such a solution
may not
exist in which case we have an anomalous gauge theory. Infinite degrees of
freedom
also  requires a regularization in the calculations. These
difficulties will not appear here since we only  consider  finite degrees of
freedom in the following.

The
condition that the original state space is an inner product space leads to
a more precise algorithm than \r{1} and provides for a direct connection to the
BFV path integral formulation. A first algorithm was given in
 \cite{Princ,Bigrad,Simple} where it was proposed that the BRST
charge operator, $Q$, must be possible to  decompose as follows (we use graded
commutators)
 \be
&&Q=\del+\del^{\dag},\;\;\;\del^2=0,\;\;\;[\del,
\del^{\dag}]\equiv\del\del^{\dag}+\del^{\dag}\del=0,
\e{2}
in which case the inner product solutions of \r{1} are the solutions of
 \be
&&\del|ph\hb=\del^{\dag}|ph\hb=0.
\e{3}
(The decomposition \r{2} requires dynamical Lagrange multipliers and
antighosts in
general.) The decomposition \r{2} and $|ph\hb$ are expected to be unique up
to BRST
invariant unitary transformations in topological trivial theories. The
solutions of
the conditions
\r{3} obtained so far  turns out to have the general form
\cite{Simple, Gauge,Solv,Geza, Bas}
\be
&&|ph\hb=e^{[Q, \psi]}|\phi\hb,
\e{4}
where $\psi$ is a hermitian, odd gauge fixing operator with ghost number minus
one, and where $|\phi\hb$ is a  BRST invariant state determined by a set of
simple hermitian operators. (The inner product of \r{4} is manifestly real.)

In a second  approach to BRST quantization on
inner product spaces the BRST invariant inner product states are from the very
beginning assumed to be of the form \r{4}. The $|\phi\hb$-states are then
required
to satisfy the following conditions \cite{Solv,Bas}
\be
&&B_i|\phi\hb=0, \quad B_i=a[Q, C_i]
\e{5}
where $a$ is a c-number, and where $B_i$ and $C_i$ are hermitian operators
 which are  in involutions. ($C_i|\phi\hb=0$ are then
allowed gauge fixing conditions
\cite{Aux,Solv}.) The index $i$ runs over one-fourth of the number of
independent
unphysical operators. (Conditions \r{5} imply then $Q|\phi\hb=0$.) However,
these
conditions are not sufficient to make the physical states \r{4} inner
product states.
\r{5} is similar to a Dirac quantization and it is well-known that an odd
$B$ leads
to a zero norm state while an even
$B$ leads to a state with infinite norm if the spectrum of $B$ is
continuous. In order
for the zeros and infinities to cancel we must have an equal number of even
and odd
$B_i$-operators. This requires the total number of $B_i$-operators to be
even. (This
is always  the case in the presence of dynamical Lagrange multipliers and
antighosts.) However, the inner product
of the $|\phi\hb$-state in \r{4} is still undefined without an appropriate
regulator factor $e^{[Q, \psi]}$. The precise condition for the hermitian gauge
fixing fermion is
\cite{Solv}
\be
&&[D'_r, (D'_s)\dagg]\;\; \mbox{\small is an invertible matrix operator}
\e{6}
where
\be
&&D'_r\equiv e^{[Q, \psi]}D_re^{-[Q, \psi]}
\e{7}
where in turn $D_r\equiv\{B_i, C_i\}$ are the hermitian BRST doublets in
involution.
Condition \r{6} implies that $D'_r$ and $(D'_r)\dagg$ form generalized BRST
quartets
\cite{KO}.

The conditions
\r{5} and
\r{3} are connected if the $\del$ operator in \r{2} is of the form
\be
&&\del =A\Big(e^{[Q, \psi]}B_ae^{-[Q, \psi]}\Big)\dagg\Big(e^{[Q,
\psi]}B_a'e^{-[Q,
\psi]}\Big),
\e{8}
where $B_a$ and $B'_a$ are the bosonic and fermionic parts of the
$B_i$-operators
($B_i=\{B_a, B'_a\}$). The conditions \r{3} yield then always \r{4} where
$|\phi\hb$
satisfies
\r{5} if $A$ is  a bosonic factor which commutes with the other
factors in $\del$.   Since the expression \r{8} does not satisfy the
 properties
\r{2} for all possible $B$-operators and all possible gauge fixings
fermions $\psi$,
the relation between the two approaches is unclear. However, for the  models
considered so far there is no  restriction  on the physics to require \r{8}
to satisfy
\r{2}.  One may notice that the choice of
$B_i$-operators in \r{5} is not unique. There are unitary equivalent sets
where the
unitary operators are BRST invariant. In addition there are different sets which
yields the same restricted $|\phi\hb$ which will be exemplified in the next
section.

 The inner
product of the states \r{4}
 will in the path integral formulation include the  form proposed by BFV,
and the
conditions on $|\phi\hb$ appear then as boundary conditions \cite{Path} as
will also
be demonstrated in  section 6. However, there are some technical points in this
correspondence which have to do with  the reality properties of the effective
Lagrangian. BRST quantization requires half of the unphysical variables to be
quantized with indefinite metric states which means that half of the basic
hermitian
unphysical operators must have imaginary eigenvalues. This implies that the
effective
Lagrangian from the operator version is not real in general. However,  it
leaves
a certain freedom in the choice of quantum states which means that it produces
several possible choices of  effective Lagrangians. There is no hint of these
choices within the path integral formulation off hand.

 When the
gauge theory contains a nontrivial Hamiltonian the above operator
quantization on
inner product states is not well defined in general. The only well defined
general
approach  seems then to be to reformulate the original theory as a
reparametrization
invariant one. The reason is that only in a reparametrization invariant
theory is the
Hamiltonian  trivial (zero) and we may rely on the previous formulation. In the
resulting framework we may then forget about the Hamiltonian and only
worry about one
nilpotent BFV-BRST charge operator now containing the Hamiltonian itself
\cite{Time}. This procedure also applies to ordinary quantum mechanics in
which case
one finds  just cohomological quantum mechanics which will be treated here.

\setcounter{equation}{0}
\section{Formal BRST solutions of cohomological quantum mechanics.}
In the introduction we reviewed how a theory described by a regular
Lagrangian $L(t)$
may be replaced by the singular reparametrization invariant Lagrangian
$L'(\tau)\equiv  L(t(\tau))\dot{t}$. As the original theory given by $L(t)$
may be  equivalently described by the first order phase space Lagrangian
$L_0(t)=p_i\dq^i-H(t)$, where $q^i$ and $p_i$ are the phase space
variables, also
the reparametrization invariant theory $L'(\tau)$ may be described by an
equivalent
 first order phase space Lagrangian. It is given by
\be
&&L_1(\tau)\equiv p_i\dq^i+\pi\dt-v(\pi+H(t)),
\e{203}
where $v$  is a Lagrange multiplier. The original regular theory is then
turned into
an abelian gauge theory with the two constraints $\pi+H(t)$ and $\pi_v=0$, where
$\pi_v$ is the conjugate momentum to the Lagrange multiplier $v$.

We   quantize now the reparametrization invariant theory $L'(\tau)$ as described
by
$L_1(\tau)$.  The original variables,
$p_i, q^i,
\pi, t,
\pi_v, v$ are then turned into operators which we denote by capital letters. The
nontrivial part of their commutator algebra is ($\hbar=1$)
\be
&&[Q^i, P_j]=i\del^i_j,\quad [T, \Pi]=i, \quad [V, \Pi_V]=i.
\e{204}
Following the BFV-prescription for the BRST charge we find the  odd,
nilpotent operator
\be
&&Q=\ca(\Pi+H(T))+\bapet\Pi_V,
\e{205}
where we have introduced the odd, hermitian ghost operators $\ca, \baca$
and their
conjugate hermitian momenta $\pet, \bapet$. ($\baca$ is an antighost.) Their
nontrivial commutators are
\be
&&[\ca, \pet]=1, \quad [\baca, \bapet]=1.
\e{206}
$H(T)$ in \r{205} is the hermitian Hamiltonian operator corresponding to the
original classical Hamiltonian in \r{102} and \r{203}. In \cite{Bas} we
made a rather
complete investigation  of simple abelian models with linear gauge fixings
in which
case it was explicitly demonstrated that
\r{2}-\r{3} lead to solutions of the form \r{4}.

 We look  now directly for
solutions of the form \r{4}.
 First we need a BRST invariant state $|\phi\hb$ determined by
linear hermitian constraints of the form $B_i|\phi\hb=0$ where $B_i=[Q, C_i]$.
In the present case we have eight unphysical operators which means that we
have two
hermitian
$B_i$-operators, one odd and one even according to the rules given in the
previous section. One allowed choice of conditions is
\be
&&\Pi_V|\phi\hb=\ca|\phi\hb=0\quad\Leftrightarrow\quad[Q,\baca]|\phi\hb=[Q,
T]|\phi\hb=0.
\e{207}
Another is
\be
&&\bapet|\phi\hb=\Big(\Pi+H(T)\Big)
|\phi\hb=0\quad\Leftrightarrow\quad[Q,\pet]|\phi\hb=[Q,
V]|\phi\hb=0.
\e{213}
Both these choices imply $Q|\phi\hb=0$.
Note, however,  that
\be
&&\ca|\phi\hb_0=\bapet|\phi\hb_0=0,
\e{208}
or
\be
&&\Pi_V|\phi\hb_0=\biggl(\Pi+H(T)\biggr)|\phi\hb_0=0,
\e{209}
are not allowed although they imply $Q|\phi\hb=0$. Eq.\r{208} makes
$|\phi\hb$ have
ghost number plus one which implies that
$|ph\hb$ in \r{4} is a zero norm state for whatever choice of $\psi$. For
\r{209} on
the other hand there exists no choice of $\psi$ such that the corresponding
$|ph\hb$
is an inner product state with finite norm. Only \r{207} and \r{213}
contain one even
and one odd
$B$-operator.  They also determine the  allowed boundary conditions in the path
integral formulation as will be seen in section 6. (Boundary conditions
corresponding
\r{207} have \eg been considered by Henneaux and Teitelboim \cite{HT,MH}.)

 In abelian gauge theories all  allowed linear conditions on $|\phi\hb$ are
related
  by means of extended unitary gauge transformations of the form \cite{Bas}
\be
&&|\phi\hb'\equiv U|\phi\hb,\quad U=\exp{\{i[Q,\rho]\}},
\e{210}
where  $\rho$ is an odd, hermitian operator with ghost number minus one.
($U$ has
then ghost number zero and does not affect the ghost number properties of the
states.) By means of
$U$ one may go to a unitary equivalent basis of operators,
\be
&&\tilde{A}=UAU^{\dag},
\e{2111}
where $A$ is any operator. In particular we have $\tilde{Q}=Q$. From \r{207} we
find then that the $|\phi\hb'$-state \r{210} satisfies
\be
&&\tilde{\Pi}_V|\phi\hb'=\tca|\phi\hb'=0,
\e{212}
if \eg $|\phi\hb$ satisfies \r{207}. This represents essentially all
allowed choices
of
$|\phi\hb$ in
\r{4} as well as all allowed boundary conditions in the BFV path integral
formalism
here. For instance,  the special conditions
\r{213} are included in \r{212}. To see this it is sufficient to consider the
following special
$U$ (an abelian subgroup of \r{210})
\be
&&U(\theta)=\exp{\{i[Q,
\rho(\theta)]\}}=\exp{\Big\{\theta\Big(V(\Pi+H(T))-
T\Pi_V+i\pet\bapet-i\baca\ca\Big)\Big\}},\nn\\
&&\rho(\theta)\equiv \theta(\pet V-\baca T),
\e{214}
where $\theta$ is a real parameter.
The conditions \r{212} become then explicitly ($|\phi\hb_{\theta}\equiv
U(\theta)|\phi\hb$)
\be
&&\Big(\Pi_V\cos\theta-(\Pi+H(T))\sin\theta\Big)|\phi\hb_{\theta}=0,\nn\\
&&\Big(\ca\cos\theta+\bapet\sin\theta\Big)|\phi\hb_{\theta}=0.
\e{215}
Thus, for $\theta=0$ we have \r{207} while for $\theta=\pi/2,\;3\pi/2$ we have
\r{213}.

Now the inner product of the  $|\phi\hb$-states \r{207} or \r{213} are not well
defined. We need a gauge fixing factor
$e^{[Q,\psi]}$ where $\psi$ is an odd, hermitian gauge fixing operator
according to
the prescription \r{4}. For the $|\phi\hb$-state \r{207} we may choose the
physical
inner product solution to be
\be
&&|ph\hb=e^{[Q,\psi]}|\phi\hb,\quad \psi=\pet\Lambda(V),
\e{216}
which is satisfactory if $\Lambda'(V)>0$ or $\Lambda'(V)<0$. The simple
argument for this is that $\Lambda(V)$ is a gauge fixing operator to
$\Pi_V$ which
annihilates $|\phi\hb$. The precise argument follows from the basic criterion
\r{6} (see below). Similarly we may choose
\be
&&|ph\hb=e^{[Q,\psi]}|\phi\hb,\quad \psi=\baca\chi(T),
\e{2151}
for the $|\phi\hb$-state \r{213}. This is satisfactory if $\dot{\chi}(T)>0$ or
$\dot{\chi}(T)<0$. There are of course more general forms allowed for $\psi$. In
particular, we may choose forms which are good for whatever choice of
conditions on
$|\phi\hb$. A natural choice is then
\be
&&\psi=\pet\Lambda(V)+\baca\chi(T),
\e{2152}
which is in accordance with the BFV prescription for path integrals. In
fact, this
$\psi$ reduces to the ones in \r{216} and \r{2151} at least  for linear
$\Lambda$ and
$\chi$. Thus, the conditions on $|\phi\hb$ makes a whole class of gauge fixing
fermions  equivalent. Other forms of $\psi$ are obtained by extended unitary
gauge transformations of the form \r{210} which does not affect the
conditions on
$|\phi\hb$ \cite{Bas}.

It is straight-forward to prove that \r{216} is an inner product solution
by means of
the basic criterion \r{6}.  The  set of hermitian BRST
doublets in involutions are here
$D_r\equiv\{\Pi_V, \ca, \chi(T), \baca\}$ where $\chi(T)$ and $\baca$ are
hermitian gauge fixing variables. (We may impose the conditions
$\chi(T)|\phi\hb=\baca|\phi\hb=0$ provided
$\dot{\chi}(T)\neq0$  since
$[Q,
\chi(T)]=\ca\dot{\chi}(T)$ must be equivalent to $\ca$ on states.)  The
corresponding transformed nonhermitian doublets defined by \r{7}  are then
explicitly
(assuming
$\acute{\chi}=\chi(\acute{T})$)
\be
&&\acute{\Pi}_V\equiv e^{[Q,
\psi]}\Pi_Ve^{-[Q,
\psi]}=\Pi_V+i(\Pi+H(T))\Lambda'(V)+\bapet\pet\Lambda''(V), \nn\\
&&\acute{\ca}\equiv e^{[Q,
\psi]}\ca e^{-[Q, \psi]}=\ca-i\bapet\Lambda'(V),\nn\\
&&\chi(\acute{T})\equiv \chi(T -i\Lambda(V) ) ,\nn\\
&&\acute{\baca}\equiv\baca +i\Lambda'(V)\pet.
\e{2161}
The only nonzero elements of the matrix operator $[\acute{D}_r,
(\acute{D}_s)\dagg]$
are then
\be
&&[\acute{\ca},(\acute{\Pi}_V)\dagg ]=2\Lambda''(V)\bapet,\quad ,
[\acute{\baca},(\acute{\Pi}_V)\dagg ]=-2\Lambda''(V)\pet,\nn\\&&
[\chi(\acute{T}),
(\acute{\Pi}_V)\dagg]=2\dot{\chi}(\acute{T})\Lambda'(V),
\quad [\acute{\ca}, (\acute{\baca})\dagg]=-i2\Lambda'(V).
\e{2162}
Hence, a necessary condition for the matrix operator $[\acute{D}_r,
(\acute{D}_s)\dagg]$ to be  nonsingular is that
$\Lambda'(V)\neq0$ as asserted above ($\dot{\chi}\neq0$).   Note that
$|ph\hb$ in
\r{216}  is  a solution to
\be
&&\acute{\Pi}_V|ph\hb=\acute{\ca}|ph\hb=0
\e{2163}
due to \r{207}.
However, the rules above also allow for the dual conditions
\be
&&(\acute{\Pi}_V)\dagg|ph'\hb=(\acute{\ca})\dagg|ph'\hb=0,
\e{2164}
with the solution
\be
&&|ph'\hb=e^{-[Q,\psi]}|\phi\hb,\quad \psi=\pet\Lambda(V).
\e{2165}
This solution is equivalent to \r{216} with $\psi=-\pet\Lambda(V)$.
The two solutions \r{216} and \r{2165} are related by
$\Lambda\leftrightarrow-\Lambda$ and correspond to the two possibilities
$\Lambda'(V)>0$ and $\Lambda'(V)<0$.

The BRST
charge
\r{205} may be decomposed according to \r{2} with $\del$ given by
\be
&&\del={i\over2\Lambda'(V)}(\acute{\Pi}_V)\dagg\acute{\ca},
\e{2167}
provided $\Lambda'(V)$ is a nonzero constant, \ie provided $\Lambda(V)$ is
linear in
$V$. (We do not know how the decomposition \r{2} looks like for
$\Lambda''(V)\neq0$.)
Since
$\acute{\ca}$ and
$(\acute{\Pi}_V)\dagg$ commute when $\Lambda''(V)=0$ the conditions \r{3}
lead to
either the
 solution \r{216} or \r{2165}.  A restriction to a linear
$\Lambda(V)$ does not affect the physical content of the theory as we shall
see.

 All other inner product solutions than those considered above are
obtained  by means of gauge transformations
 represented by unitary operators of the form \r{210}.
 These solutions
are
\be
&&|ph\hb'=U|ph\hb=e^{\pm[Q,\tpsi]}|\phi\hb', \quad \tpsi=U\psi U^{\dag},
\e{217}
where $|\phi\hb'$ is given by \r{210}. That $U$ generates gauge
transformations on
$|ph\hb$ follows provided $|ph\hb$
is an inner product state since $|ph\hb'=U|ph\hb=|ph\hb+Q|\;\hb$. Notice also
that there are gauge transformations that change
$\psi$ but not the conditions
\r{207} (or the opposite way around as will be shown below). For instance,  the
results in \cite{Bas} suggests that we may  have the representation
\r{216} with $\psi=\pet\Lambda(V)+\pet\mu(\Pi+H(T))$ for arbitrary
parameter $\mu$.

Since the gauge theory we are considering does not contain any
topological obstructions
it is natural to expect \r{217} to represent all possible solutions for one
definite
sign in the exponential. However, this is {\em not} the case. It is impossible
to get $-\psi$ by a unitary gauge transformation of
$\psi$. Thus, there are two disconnected sets of solutions even in
topologically trivial theories.
 Either there is a condition which excludes one of
these sets or they are equivalent. In the next section  we  find  that for
cohomological quantum mechanics one set has negative norms and must be
excluded.
(This is a general property  when we have an odd number of original
constraints
\cite{Bas}.)

The BRST invariant states \r{217}  are defined up to zero norm states. In
order to determine unique states representing the original theory we have
to project
out the BRST singlets \cite{KO}, which we denote by $|s\hb$. The BRST
singlets have exactly the same form as
 \r{217}  except that the $|\phi\hb$-states satisfy two extra gauge fixing
conditions. The allowed class of gauge fixing conditions was specified in
\cite{Solv}.
For \r{216} and \r{2165} the natural choice is  \cite{Time}
\be
&&\chi(T)|\phi\hb=\baca|\phi\hb=0,
\e{218}
where $\chi(T)$ is a hermitian gauge fixing to $\Pi+H(T)$.
($\dot{\chi}(T)\neq0$ is
required.) The last condition in \r{218} completely ghost fixes $|\phi\hb$ and
makes the corresponding singlet $|s\hb$ have ghost number zero. For \r{213} the
natural gauge fixing conditions are
\be
&&\Lambda(V)|\phi\hb=\pet|\phi\hb=0,
\e{219}
where $\Lambda'(V)\neq0$ is required. In the following it will be shown
that linear
$\chi$ and $\Lambda$ are always allowed and that a restriction to linear
$\chi$ and
$\Lambda$ have no physical consequence.

 All the solutions given here are formal. To actually determine true inner
product
solutions we must specify the representation of the original extended state
space.
When this is done we will be able to give wave function representations of
the BRST
singlets
$|s\hb$.

\setcounter{equation}{0}
\section{Wave function representations}
The freedom in the choice of an extended state space in a BRST theory has to be
investigated for each model under consideration. For cohomological quantum
mechanics
we assume  the original theory to be  regular and possible to quantize. It is
therefore natural to assume the operators
$P_i$, $Q^i$  to span a Hilbert space, in which case $P_i$, $Q^i$ must have real
eigenvalues on appropriate eigenstates. However, time $T$ and the Lagrange
multiplier
$V$ as well as the ghosts belong to the unphysical part of the state space. This
means that half of these variables must span an indefinite metric state
space. The
corresponding half of the hermitian operators have then imaginary
eigenvalues on the
appropriate eigenstates. Now exactly which half of the unphysical operators have
these properties is not specified off-hand. The different choices for the
fermions
are simply related by phase factors like signs or factors of $i$, but this
is not
the case for the bosons. The ambiguity is vary large. $T$ may \eg have arbitrary
complex eigenvalues
\cite{Bas}. Now even the separation into physical and unphysical variables
are not
that simple either. In fact, physical operators should be BRST invariant,
and $P_i$
and $Q^i$ are not. However, we know there exist BRST invariant operators
representing
the $P_i$- and $Q^i$-operators of the  original theory. (These so called gauge
invariant extended operators are expected  to be closely related to
$P_i$ and
$Q^i$ and to have real eigenvalues (cf appendix A).) Below we let the
possible wave
function representations determine the possible choices of an extended
state space.

Consider the particular BRST singlet
\be
&&|s\hb=e^{[Q, \psi]}|\phi\hb,\quad \psi=\pet\Lambda({V}),
\e{31}
 where $|\phi\hb$ satisfies the
conditions \r{207} and \r{218}.
We consider first the wave function representations of the $|\phi\hb$-states.
The conditions
\r{207} and \r{218} ghost fix $|\phi\hb$ to
\be
&&|\phi\hb=|\phi\hb_B\otimes|0\hb_{\ca\baca},
\e{301}
where $|0\hb_{\ca\baca}$ is a ghost vacuum, and where the bosonic part
$|\phi\hb_B$
satisfies
\be
&&\Pi_V|\phi\hb_B=0, \quad \chi(T)|\phi\hb_B=0.
\e{302}
The formal wave function representation of the solutions is
\be
&&\vb   q^i, t, v,  \pet,\bapet|\phi\hb=\del(\chi(t))\varphi(q, t),
\e{303}
where $q^i$, $t$, $v$, $\pet$, $\bapet$  are eigenvalues of the
operators ${Q}^i$,  $T$,
${V}$, $\pet$, $\bapet$   respectively. (One may of course also
consider wave functions depending on the eigenvalues of the ghosts $\ca, \baca$
instead of $\pet, \bapet$. However, in this case the right hand side would
involve the delta function
$\del(\ca)\del(\baca)$ as an additional factor. We find the
representation
\r{303} more convenient to use here.) Now a solution of the type \r{303} is
only consistent if $\chi(t)$ is real up to a constant phase factor which
restricts
the representation and/or $\chi(t)$.   Furthermore, we must
require $\chi(t)=0$ to have a unique solution, $t=t_0$, which requires
$\chi(t)$ to be a monotonic function of $t$. (Otherwise, we would have gauge
inequivalent sectors and we would not be able to obtain the original
theory.) This
means that $\chi(t)$ always may be replaced by the linear expression
$t-t_0$.  Notice
that the inner product of
\r{303} leads to two zero factors from the
$\pet$ and
$\bapet$ integrations, and to two infinite factors from the
$v$ and $t$ integrations. ($\dot{\chi}(t_0)$ is assumed to be a finite  and
$\varphi$ is assumed to be such that $\int d^nq \varphi^*\varphi<\infty$.)

 What is the wave function representation of
 the  singlet \r{31}? The
conditions
\r{207} and
\r{218} on
$|\phi\hb$ imply the following conditions on $|s\hb$ using the gauge fixing
fermion \r{31} (cf. \r{2161} and the treatment in
\cite{Time})
\be
&&\biggl(\Pi_V+i{\Lambda}'(V)\left(\Pi+H(T)\right)+
{\Lambda''(V)}\bapet\pet\biggr)|s\hb=0\nn\\
&&\biggl(\ca-i{\Lambda'(V)}\bapet\biggr)|s\hb=\biggl(\baca+
i{\Lambda'(V)}\pet\biggr)|s\hb=0\nn\\
&&\chi(T-i{\Lambda(V)})|s\hb=0,
\e{305}
which yield the following solution in the wave function representation
\be
&&\Psi(q, t,v,\pet,\bapet)\equiv\vb q^i, t,v,\pet,\bapet
|s\hb=\nn\\&&=e^{-i\Lambda'(v)\bapet\pet}
\del(\chi(t-i\Lambda(v)))
e^{\Lambda(v)(-i\partial_{t}+H_S(t))}\varphi(q, t),
\e{306}
where
$\varphi(q, t)$ is the same wave function as in
\r{303}. $H_S(t)$ is the Schr\"odinger
representation of the operator
$H(T)$ defined by
\be
&&H_S(t)\vb  q^i, t, v,  \pet,\bapet|\phi\hb\equiv\vb  q^i, t, v,
\pet,\bapet|H(T)|\phi\hb.
\e{3062}
The differential operator in \r{306} may be worked out explicitly (see
\r{3151} and \r{3103} below).

The representations \r{303} and \r{306} are formal as long as we do not
specify the
properties of the eigenvalues. A precise choice of state space determines these
eigenvalues. Since the argument of the delta function in
\r{306}
 must be real up to a constant phase
factor, we arrive at the following
 two natural options what regards the bosonic variables:

Case 1: $t$ is real, $v$ is imaginary, and $\Lambda(v)$ is imaginary

Case 2: $t$ is imaginary, $v$ is real, and $\chi(t)$ is imaginary

(One may also choose complex eigenvalues  which correspond to
representations between these two possibilities \cite{Bas}.) For the fermions we
choose $\pet$ real and $\bapet$ imaginary ($=i\bapet$, $\bapet$ real). Some
properties of the corresponding ghost states are given in appendix B.

Consider case 1
first. Put $v=iu$ where $u$ is real and assume $\chi(t)=0$ to have the unique
solution
$t=t_0$.   The eigenstates of
$V$ have  the properties
\cite{HT,Gen}
\be
&&V|iu\hb=iu|iu\hb,\;\;\;\vb
-iu|\equiv(|iu\hb)^{\dag},\;\;\;\vb iu|iu'\hb=\del(u-u')\nn\\ &&\int du|iu\hb\vb
iu|=\int d u|-iu\hb\vb -iu|=\bett.
\e{3061}
We find then from \r{306} ($\pet$, $\bapet$ real)
\be
&&\vb s|s\hb=\int d^nq dt d u  d\pet d\bapet\, \Psi^*(q, t, -iu, \pet, -i\bapet
)\Psi(q, t, iu, \pet, i\bapet)=\nn\\
&&=-\sign(\Lambda'(0))\int d^n q {1\over |\dot{\chi}(t_0)|^2}\varphi^*(q,
t_0)\varphi(q, t_0)
\e{307}
provided $i\Lambda(iu)$ is a real monotonic function. Since $\Lambda$ is real
for real argument, $\Lambda(iu)$ must be an odd function of $u$ in
order to be imaginary. Hence, if $i\Lambda(iu)$ is monotonic in $u$ then
$\Lambda(iu)=0$ has the unique solution $u=0$. (If $i\Lambda(iu)$ is not
monotonic
and $\Lambda(iu)=0$ has several solutions then \r{307} is replaced by a sum
of terms
which is not what we should have.) Obviously a restriction to a linear
$\Lambda(iu)$
has no physical consequence.

 The expression \r{307} implies that $\vb s|s\hb$ is only positive
definite for $\Lambda'<0$. (Notice that $\Lambda'(iu)$ is  real.) This
means that the
two classes of gauge fixing characterized by opposite signs of $\psi$ have
opposite
norms here. (This is in general the case if the number of original
constraints is odd
\cite{Bas}.) Thus, since we have positive norms for $\Lambda'<0$ we have to
exclude
the choice
$\Lambda'>0$. The unitary gauge transformations can obviously not change
the sign of
$\Lambda'$.

Consider now case 2. For $t=iu$,  $i\chi(iu)$
 must be real and monotonic in $u$. Since $\chi$ is real
for real argument, $i\chi(iu)$ must be  odd in $u$.
Thus, $\chi(iu)=0$ has the unique  solution $u=0$.  (A
linear
$\chi$ is always possible.) We find now with the same conventions as above
\be
&&\vb s|s\hb=\int d^nq d u dv   d\pet d\bapet\, \Psi^*(q, -iu, v, \pet, -i\bapet
)\Psi(q, iu, v, \pet, i\bapet)=\nn\\
&&=-\sign(\Lambda'(v_0))\int d^n q {1\over |\dot{\chi}(0)|^2}\varphi^*(q,
0)\varphi(q,
0),
\e{311}
provided $\Lambda(v)$ is monotonic such that $\Lambda(v)=0$ yields the
unique value
$v=v_0$. (If this is not the case we get a sum in \r{311} which we should
not have.)
Also
\r{311} is  positive definite  for $\Lambda'<0$. This is  a particular case of
\r{307}, since  $\chi$ here is restricted and does not yield any explicit
parameter
$t_0$.

The only parameters the gauge fixing condition $\chi(T)|\phi\hb=0$
introduces are
$t_0$ and $\dot{\chi}$. Since the inner products
of the BRST singlets $|s\hb$ cannot depend on these gauge  parameters the
results above suggest that we should define the physical wave function by
($t_0=0$ in
case 2)
\be
&&\phi(q,t_0)\equiv {1\over |\dot{\chi}(t_0)|}\varphi(q,t_0),
\e{308}
since \r{307} then yields
\be
&&\vb s|s\hb=\int d^nq \,\phi^*(q, t_0)\phi(q, t_0),
\e{309}
for $\Lambda'<0$. Obviously $\phi$ is independent of $\dot{\chi}$ since it is
$\phi$ that is normalized by $|s\hb$. The BRST quantization requires
\r{309} to be
independent of the gauge  parameter $t_0$. It is therefore natural to expect
that different values of
$t_0$ may be reached by
unitary gauge transformation of the form \r{210}. An appropriate unitary gauge
operator is
\be
&&U(a)\equiv
e^{i[Q,\rho]}=e^{-ia(\Pi+H(T))},
\quad
\rho\equiv -a\pet,
\e{312}
where $a$ is a real constant. This $U(a)$ does not alter the gauge fixing
fermion
$\psi$ in \r{31} which means that
\be
&&|s'\hb=U(a)|s\hb=e^{[Q, \psi]}|\phi'\hb, \quad |\phi'\hb=U(a)|\phi\hb,
\e{3121}
where $|\phi'\hb$ satisfies the same
 conditions as $|\phi\hb$ except for the gauge fixing
condition which becomes
\be
&&\chi(T-a)|\phi'\hb=0,
\e{313}
since $U(a)\chi(T)U^{-1}(a)=\chi(T-a)$.
Thus, $U(a)$ induces  translations in $t$. We get therefore
\be
&&\vb q^i,  t, v, \pet,\bapet
|\phi'\hb=\del(\chi(t-a))\varphi'(q,t)\equiv\del(t-t_0-a)\phi'(q, t),\nn\\
&&\vb
 q^i,  t, v,
\pet,\bapet|\phi\hb=\del(\chi(t))\varphi(q,t)\equiv\del(t-t_0)\phi(q, t),
\e{315}
where
\be
&&\phi'(q, t)={1\over |\dot{\chi}(t_0)|}\varphi'(q,t).
\e{314}
The
relation
$|\phi'\hb=U(a)|\phi\hb$ implies
\be
&&\left.\phi'(q, t_0+a)=e^{-a(\dif_{t}+iH_S(t))}\phi(q,
t_0)\right|_{t=t_0+a}.
\e{3151}
This means that $\phi'(q, t)$ satisfies the Schr\"odinger equation with
respect to
$t$ and that
$\phi'(q, t_0)=\phi(q, t_0)$. The relation \r{3121} implies furthermore that
\be
&&\vb s|s\hb=\vb s'|s'\hb=\int d^nq \,\phi^{\prime *}(q, t)\phi'(q, t).
\e{3152}
The wave function $\phi'(q, t)$ seems therefore to be a more appropriate
physical
wave function than $\phi(q, t_0)$ since it satisfies the Schr\"odinger
equation.
If we view $\phi'(q, t)$ as the physical wave function at arbitrary times then
$\phi(q, t_0)$ is the physical wave function at $t=t_0$ since $\phi'(q,
t_0)=\phi(q,
t_0)$. However, a peculiar property of $\phi'(q, t)$ is that it depends on the
gauge  parameter
$t_0$ on which no physical property should depend.   On the other hand this is
expected if we view $\phi'(q, t)$ as  the
gauge invariant extension of $\phi(q,t_0)$:   That it satisfies the
Schr\"odinger
equation expresses its gauge invariance, and that  $\phi'(q, t_0)=\phi(q,
t_0)$   tells us
that $\phi'(q, t)$ indeed is the gauge invariant extension of the wave function
$\phi(q, t_0)$.  (Compare the properties of the
classical gauge invariant extensions considered in appendix A. There we have
  $-ad({H}(t))$ instead of $iH_S(t)$.)

 Eq.\r{3151} may be rewritten as follows
\be
&&\phi'(q, t)=U(t,t_0)\phi(q, t_0),
\e{3103}
where
\be
&&U(t, t_0)=\left\{\begin{array}{ll}{\cal T}\exp{\{-i\int_{t_0}^tdt'H_S(t')\}},
&t>t_0\\
\tilde{\cal T}\exp{\{-i\int_{t_0}^tdt'H_S(t')\}}, &t<t_0,
\end{array}\right.
\e{3104}
where ${\cal T}$ and $\tilde{\cal T}$ denote time and antitime ordering
respectively.
Since $U(t, t_0)$ is a unitary operator \r{3103} confirms the relation
\r{3152}.
  We notice that if $\phi(q, t_0)$  satisfies the
Schr\"odinger equation, then $\phi'(q, t)$   is independent
of
$t_0$ due to \r{3103} and we have
$\phi'(q, t)=\phi(q, t)$. This  could therefore to be a reasonable condition to
impose.

In case 2  the parameter $a$ in $U(a)$ \r{312} must be imaginary which
implies that
$U(a)$ no longer is unitary, and that $\chi(T-a)$  no longer is hermitian.
However,  this is consistent! \r{3121} is still a singlet state. Instead of
\r{3151} we get  that $\phi'(q, iu)$ satisfies the Schr\"odinger equation with
imaginary time. However, since
\r{313}  implies
\be
&&\vb\phi|\chi(T+a)=0
\e{316}
we get
\be
&&\vb s|s\hb=\vb s|U\dagg(a)U(a)|s\hb=\int d^nq \,\phi^{\prime *}(q, 0)\phi'(q,
0)=\int d^nq
\,\phi^{ *}(q,0)\phi(q,0).
\e{3161}
(The last equality is trivial since  $\phi'(q, 0)=\phi(q,0)$.) $\phi'(q,
iu)$ for
nonzero
$u$ does therefore not represent the singlets $|s\hb$. An appropriate inner
product
for
$\phi'(q, iu)$ would otherwise be
$\int d^nq \,\phi^{\prime *}(q, -iu)\phi'(q, iu)$ which is positive
definite only if
$\phi'(q, iu)$ is even in $u$.

\subsection{Physical wave functions as projections of BRST singlets}
Since the singlet states represent the physical states, they should contain
the wave
functions of the original theory. We propose the following general
projection from
singlets to physical wave functions  depending on   the original
coordinates and a parameter which possibly could represent time
\beq
Impose  boundary conditions determined by the conditions on $|\phi\hb$ on
the wave
function representation $\Psi$ of  the BRST singlets $|s\hb$. \hfill(A)
\eq
Symbolically this may be stated as follows
\be
&&\left.\phi_{phys}(q,t_0)=\Psi\right|_{b_i=c_i=0},
\e{3101}
where $b_i$ and $c_i$ are eigenvalues or Weyl symbols of the operators
$B_i$ and $C_i$
in the defining conditions $B_i|\phi\hb=  C_i|\phi\hb=0 $. Here we have
$B_i=\{\Pi_V, \ca\}$ and
$C_i=\{\baca, \chi(T)\}$ which in case 1 yields the boundary conditions
$\pi_u=\ca=0$
and
$\baca=\chi(t)=0$. Hence, we find from \r{3101} and \r{306} ($\pet$ and
$\bapet$
are real and
$\la(u)=i\Lambda(iu)$ is a real monotonic function of $u$.
$\la'(u)=-\Lambda'>0$)
\be
&&\phi_{phys}(q,t_0)\equiv \int d u d\pet d\bapet\, \Psi(q, t_0, iu, \pet,
i\bapet)=\phi(q,t_0),
\e{3102}
\ie exactly \r{308} which we derived in another way above. Notice that
$\chi(t)=0$ is
equivalent to $t-t_0=0$, and that $\ca=\baca=\pi_u=0$ is equivalent to an
integration over $\pet, \bapet$ and $u$ due to Fourier transformations.
In case 2 we get similarly $\phi(q,0)$ as suggested by \r{311}.

There is also an
interesting partial projection which may be stated as follows:
\beq
Impose the conditions on $|\phi\hb$ except for the gauge fixing conditions as
boundary conditions on the wave function representation of the BRST singlets
$|s\hb$.\hfill(B)
\eq
The $C_i$ operators contain both ghost fixings and gauge fixings of the original
invariances. The prescription B requires us not to impose the last ones as
boundary
conditions on $\Psi$.
 In
case 1 the prescription B requires us  here to put $\ca, \baca$ and $\pi_u$
equal
to zero but not $t=t_0$.  We get therefore from \r{306}.
\be
&&\left.\bar{\phi}(q,t,t_0)\equiv \int d u d\pet d\bapet\, \Psi(q, t, iu, \pet,
i\bapet
)=e^{-\la(\dif_{t}+iH_S(t))}\phi(q,t)\right|_{\la=t-t_0},
\e{310}
which means that
\be
&&\bar{\phi}(q,t,t_0)=\phi'(q,t).
\e{31001}
Thus, $\bar{\phi}(q,t,t_0)$ is the appropriate physical wave function which
satisfies
the Schr\"odinger equation.

In case 2 the partially projected wave function is given by ($\Lambda'<0$)
 \be
&&\bar{\phi}(q,iu)\equiv \int d v d\pet d\bapet\, \Psi(q, iu, v, \pet,
i\bapet
)=\left.{1\over
|\dot{\chi}(0)|}e^{-\la(\partial_u-H_S(iu))}\varphi(q,0)\right|_{\la=u},
\e{3105}
which satisfies the Schr\"odinger equation for imaginary time, $iu$. Obviously
$\bar{\phi}(q,iu)={\phi}'(q,iu)$. If we also impose the gauge fixing
condition $u=0$
we get $\bar{\phi}(q,0)={\phi}'(q,0)={\phi}(q,0)$.

It is remarkable that the physical wave function which follows from the
projection A
above
also is  directly related to the wave function representation of the
$|\phi\hb$-state according to
\r{303}. We have
\be
&&\vb  q^i,  t, v, \pet,\bapet|\phi\hb=\del(t-t_0)\phi(q, t).
\e{31042}
From this result we propose the following general projection prescription:
\beq
The physical wave functions may be obtained from the wave function of the
$|\phi\hb$-state by imposing the dual conditions to the gauge fixing
conditions on $|\phi\hb$  as boundary conditions.\hfill(C)
\eq
By a dual condition we mean the canonical conjugate condition or more
precisely the
corresponding gauge invariance.  In case 1 the dual to the gauge fixing
condition,
$\chi=0$, is as boundary condition
$\pi+H=0$. This condition may roughly be replaced by
an integration over time $t$ which obviously leads to the physical wave
function.
Notice that we can only obtain a gauge fixed physical wave function from the
$|\phi\hb$-state. (The total set of boundary conditions satisfied by $\vb
q^i,  t, v,
\pet,\bapet|\phi\hb$ corresponds to the dual set \r{213} and \r{219} of allowed
conditions on $|\phi\hb$, which also will be considered in the  next section.)

It is only projection B that leads to physical wave functions satisfying
the basic
gauge invariance which here is the Schr\"odinger equation.

\subsection{Projected physical propagators}
Consider the gauge transformed singlet
\be
&&|\tilde{s}\hb=U(\al)|s\hb, \quad U(\al)=e^{i\al[Q,\rho]},\quad\rho\equiv
\pet (T-t_0),
\e{401}
where $\alpha$ is a real parameter. With $|s\hb$ given by \r{31} we find
\be
&&|\tilde{s}\hb=e^{{\ga\over2}[Q,\psi]}|\phi\hb,
\e{402}
where $\ga$ is a positive constant which depends on $\al$ in \r{401}. Note that
$U(\al)|\phi\hb\equiv|\phi\hb$. We have therefore
\be
&&\vb s|s\hb=\vb \tilde{s}|\tilde{s}\hb=\vb
\tilde{s}|U\dagg(b)U(a)|\tilde{s}\hb=\vb\phi''|e^{\ga[Q,\psi]}|\phi'\hb,
\e{403}
where $U(a)$ and $U(b)$ are given by \r{312} and where $|\phi'\hb$ and
$|\phi''\hb$
are defined by
\r{3121}. This inner product may be calculated in the representation 1
where $t$ is
real and
$v$ is imaginary. We find then
\be
&&\vb s|s\hb=\vb\phi''|e^{\ga[Q,\psi]}|\phi'\hb=\int d^{n+4}R''
d^{n+4}R'\vb\phi|R^{\prime\prime *}\hb\vb R''| e^{\ga[Q, \psi]} |R^{\prime *}\hb\vb
R^{\prime}|\phi'\hb,
\e{405}
where $R\equiv\{q^i, t, iu, \pet, i\bapet\}$. From \r{303} we have
\be
&&\vb R^{\prime\prime *}|\phi''\hb=\del(\chi(t''))\varphi(q'', t''),\quad
\vb R^{\prime}|\phi'\hb=\del(\chi(t'-a))\varphi'(q', t').
\e{406}
Hence, we get
\be
&&\vb s|s\hb=\vb\phi''|e^{\ga[Q,\psi]}|\phi'\hb=\int d^nq'' d^nq'
\phi^{\prime\prime
*}(q'', t_0+b) K(q'',q';t_0+b,t_0+a)\phi'({q'}, t_0+a),\nn\\
\e{407}
where
\be
&&K(q'',q';t'',t')\equiv \int du'' d\pet'' d\bapet'' du' d\pet' d\bapet'
\vb q^{\prime\prime i}, t'',
iu'',
\pet'',
i\bapet''|e^{\ga[Q, \psi]} | {q}^{\prime i}, t', iu', \pet',
i\bapet'  \hb.\nn\\
\e{408}
Since \r{407} also may be written as
\be
&\vb s|s\hb&=\vb\phi''|e^{\ga[Q,\psi]}|\phi'\hb=\nn\\&&=\int d^nq'' d^nq'
\bar{\phi}^{*}(q'', t_0+b,t_0) K(q'',q';t_0+b,t_0+a)\bar{\phi}({q'}, t_0+a,t_0),
\e{4081}
where $\bar{\phi}({q'}, t,t_0)$ is the physical wave function, $K$ in
\r{408} must
be  the physical propagator.
Note that the expression \r{408} involves exactly the same integrations as
in the
formula for the partially projected physical  wave function $\bar{\phi}({q'},
t,t_0)$ given in \r{310}. Thus, we have arrived at the following general
prescription
for physical propagators:
\beq
Impose as boundary conditions on the extended propagator  $\vb
R^{\prime\prime }|e^{\ga[Q, \psi]}|R^{\prime *}\hb$ the  conditions on
$|\phi\hb$
except for the gauge fixing conditions.\hfill(D)
\eq
(In principle we may here also impose the gauge fixing conditions, but then
they should be different on the left and right-hand side which  effectively
is what
the
$|\phi\hb$-states \r{406} do in \r{407}.) The same prescription is also
valid for
 the standard BFV path integral propagators as will be shown in section
6.

We may explicitly check that \r{408} is the physical propagator by
performing the
integrations. We find
\be
&&K(q,q';t,t')=-\ga\int_{-\infty}^{\infty} du
\Lambda'(iu)e^{\ga\Lambda(iu)(-i\dif_{t}+H_S(t))}\del^n(q-q')
\del(t-t')=\nn\\
&&=\varepsilon\int_{-\infty}^{\infty} d\la
e^{-\varepsilon\la(\dif_{t}+iH_S(t))}\del^n(q-q')\del(t-t')=\varepsilon\vb q^i,
t|{q}^{\prime i}, t'\hb, \nn\\&&
\vb q^i, t|{q}^{\prime i},
t'\hb\equiv
\left.e^{-\la(\dif_{t}+iH_S(t))}\del^n(q-q')\right|_{\la=t-t'},\quad
\varepsilon=-\sign(\ga\Lambda').
\e{409}
The change of variable $u \ra \la$ in the second equality is allowed  since
$\la(u)\equiv
\pm\ga i\Lambda(iu)$ is a real monotonic function of
$u$. The sign is chosen such that
$\la'(u)>0$. Eq.\r{409} is the correct physical propagator for positive
norm states when $\Lambda'<0$ since $\ga>0$. (For $\ga<0$ $\Lambda'>0$
yields the
correct physical propagator.) Its gauge invariance follows from the last
equality in
\r{409}. We have
\be
&&\vb q^i, t|(\Pi+H(T))|{q}^{\prime i},
t'\hb=(-i\dif_{t}+H_S(t))\vb q^i, t|{q}^{\prime i},t'\hb
=0.
\e{410}
Propagators may be expressed in terms of path integrals (see section 6),
and that path
integrals could satisfy the original constraints was first proposed for the wave
function of the universe in \cite{HHw}. In \cite{HH} (see also \cite{HT})
this was
shown to be a general feature of BFV path integrals if one imposes the
generalization
of the boundary conditions considered in this section (given in \cite{MH}).
These
boundary conditions are the simplest allowed conditions one may impose and
seems also
special for these properties. However, our results of section 6
suggest that these properties should also be the valid for all allowed
boundary conditions provided an appropriate gauge fixing fermion is chosen.

That the expression \r{407} is independent of $a$ and $b$ follows since
\be
&&\bar{\phi}(q, t, t_0)=\int d^nq^{\prime i}\vb {q}^i,
t|{q}^{\prime i}, t'\hb\bar{\phi}(q', t', t_0).
\e{412}

In case 2 we cannot derive the physical propagator along the lines followed in
case 1 since the singlets are not represented by $\phi'(q,iu)$ for nonzero $u$.
However, we may still define the physical propagator by the prescription D.
We find
then
\be
&&K(q,q';iu, iu')\equiv \int dv d\pet d\bapet dv' d\pet' d\bapet' \vb q^i,
iu,v,
\pet,
i\bapet|e^{\ga[Q, \psi]} | {q}^{\prime i}, iu', v', \pet',
i\bapet'  \hb=\nn\\
&&=\varepsilon\int_{-\infty}^{\infty} d\la
e^{-\la(\dif_{u}-H_S(iu))}\del^n(q-q')\del(u-u')=\varepsilon\vb q^i,
iu|{q}^{\prime i},
iu'\hb,\nn\\
&&\vb q^i, iu|{q}^{\prime i},
iu'\hb=
\left.e^{-\la(\dif_{u}-H_S(iu))}\del^n(q-q')\right|_{\la=u'-u},
\varepsilon\equiv-\sign(\ga\Lambda'),
\e{417}
which indeed is the appropriate propagator for imaginary time and
$\ga\Lambda'<0$.

These projected propagators will also be considered within the path integral
formulation in section 6.

\setcounter{equation}{0}
\section{Representations of other BRST singlets}
So far we have considered the particular BRST singlet \r{216}. All other BRST
singlets $|\tilde{s}\hb$ are obtained from \r{216} by means of extended
unitary gauge
transformations,
\be
&&|\tilde{s}\hb=U|s\hb, \quad U=\exp{\{i[Q, \rho]\}},
\e{601} where $\rho$ is an odd, hermitian operator with ghost number one.
This implies
\be
&&|\tilde{s}\hb=e^{[Q, \tpsi]}|\tilde{\phi}\hb,
\e{602}
where $\tpsi=\tpet\Lambda(\tilde{V})$, and where $|\tilde{\phi}\hb$
satisfies the
conditions
\be
&&\tilde{\Pi}_V|\phi\hb=\tca|\phi\hb=0,\nn\\
&&\chi(\tilde{T})|\phi\hb=\tbaca|\phi\hb=0.
\e{603}
Note that all tilde operators are defined by \r{2111},
\ie $\tilde{A}\equiv UAU\dagg$,
where
$U$ is an extended gauge transformation of the form \r{601}.  The two
different sets
in
\r{217} are  contained in
\r{602} as long as we do not specify
$\Lambda(\tilde{V})$. (We have either $\Lambda'(\tilde{V})>0$ or
$\Lambda'(\tilde{V})<0$.)  Since the BRST charge \r{205}  may be
written as
\be
&&Q=\tca(\tilde{\Pi}+\tilde{H}(\tilde{T}))+\tbapet\tilde{\Pi}_V,
\e{604}
where $\tilde{H}=UHU\dagg$, we may view the tilde operators as the basic
ones. This
shows that we should get exactly the same results as above if we make use
of tilde
operators. However, in terms of the original variables the situation is
different.
Since $|\tilde{s}\hb$ is different from $|s\hb$ their wave function
representations
should  be different. Below we treat therefore another BRST singlet in detail.

We consider here the particular BRST singlets
\be
&&|s\hb=e^{[Q, \psi]}|\phi\hb, \quad \psi=\baca\chi(T)
\e{a1}
where $|\phi\hb$ satisfies the conditions \r{213} and \r{219}, \ie
\be
&&\bapet|\phi\hb=\left(\Pi+H(T)\right)|\phi\hb=0,\nn\\
&&\Lambda(V)|\phi\hb=\pet|\phi\hb=0,
\e{a2}
which are dual to \r{207} and \r{218}.
These singlets are contained in the singlets \r{602}. Take \eg the gauge
transformation \r{214} with $\theta=\pi/2,\;3\pi/2$. The wave function
representation
of the $|\phi\hb$-state satisfying \r{a2} is
\be
&&\vb  q^i, t, v,  \ca,\baca|\phi\hb=\del(\Lambda(v))\varphi(q, t, v),
\e{a3}
where $\varphi(q, t, v)$ satisfies the Schr\"odinger equation with respect
to $t$.
This is quite different from \r{303} which did not involve any Schr\"odinger
equation. Again $\Lambda(v)=0$ is required to yield the unique solution
$v=v_0$ which
implies that we always may choose $\Lambda(v)$ to be linear in $v$.  The
conditions
\r{a2} imply the following conditions on
$|s\hb$ in
\r{a1}:
\be
&&\biggl(\Pi+H(T)
+i\dot{\chi}(T)\Pi_V+\ddot{\chi}(T)\ca\baca\biggr)|s\hb=0\nn\\
&&\biggl(\pet+i\dot{\chi}(T)\baca\biggr)|s\hb=\biggl(\bapet-
i\dot{\chi}(T)\ca\biggr)|s\hb=0\nn\\ &&\Lambda(V-i{\chi}(T))|s\hb=0,
\e{a4}
which yield the following solution in the wave function representation
\be
&&\Psi(q, t,v,\ca,\baca)\equiv\vb q^i, t,v,\ca,\baca
|s\hb=\nn\\&&=e^{i\dot{\chi}(t)\baca\ca}
\del(\Lambda(v-i\chi(t))
\varphi(q, t, v-i\chi(t)),
\e{a5}
where
$\varphi(q, t, v)$ is the same wave function as in
\r{a3}.

Let us calculate the inner product for the following
 two natural representations:

Case a: $t$  imaginary, $v$  real, and $\chi(t)$
imaginary. $\ca$ real and $\baca$
imaginary ($=-i\baca$, $\baca$ real).

Case b: $t$  real, $v$  imaginary, and $\Lambda(v)$
 imaginary. $\ca$ real and $\baca$
imaginary ($=-i\baca$, $\baca$ real).

(By means of the gauge transformations \r{214}
 with $\theta=\pi/2,\;3\pi/2$ they
correspond  to the two cases considered in section 4.)

 Following the treatment
in section 4 we find here in case a,
\be
&&\vb s|s\hb=\int  d u d v d^nq  d\baca d\ca
\,\Psi^*(q, -iu, v, \ca, i\baca
)\,\Psi(q, iu, v,  \ca, -i\baca)=\nn\\
&&=\sign(\dot{\chi}(0))\int d^n q {1\over |\Lambda'(v_0)|^2}\varphi^*(q, 0,
v_0)\varphi(q, 0, v_0),
\e{a6}
provided $i\chi(iu)$ and $\Lambda(v)$ are monotonic functions.
 In case b we find,
\be
&&\vb s|s\hb=\int dt d u   d^nq  d\baca d\ca
\,\Psi^*(q, t, -iu, \ca, i\baca
)\Psi(q, t, iu,  \ca, -i\baca)=\nn\\
&&=\sign(\dot{\chi}(t_0))\int d^n q {1\over |\Lambda'(0)|^2}\varphi^*(q,t_0,
0)\varphi(q, t_0, 0),
\e{a7}
provided $\chi(t)$ and $i\Lambda(iu)$ are monotonic functions.
Positive norms requires
 $\dot{\chi}>0$ in both cases. Since
$\varphi(q, t_0, 0)$ satisfies the Schr\"odinger
equation with respect to $t_0$ we get in  case b
\be
&&\vb s|s\hb=\sign(\dot{\chi}(t_0))\int d^n q {1\over
|\Lambda'(0)|^2}\varphi^*(q,0,
0)\varphi(q, 0, 0).
\e{a8}
We may now compare \r{a6} and \r{a8} with \r{307} and \r{311}. We find then
 agreement if we replace $\Lambda(v)$ by $\chi(-t)$ or $-\chi(t)$ as required
by the gauge transformation \r{214} with $\theta=\pi/2,\;3\pi/2$. We should
also exchange $v_0$ and $t_0$.

From \r{a6} we may identify a physical wave function in case a to be
given by ($\dot{\chi}>0$)
\be
&&\phi_{(a)}(q,v_0)\equiv{1\over|\Lambda'(v_0)|}\varphi(q,0,v_0),
\e{a9}
and in case b ($\dot{\chi}>0$)
\be
&&\phi_{(b)}(q,t_0)\equiv{1\over|\Lambda'(0)|}\varphi(q,t_0,0).
\e{a10}
$\phi_{(a)}$ does not satisfy the Schr\"odinger equation with respect to
$v_0$ as in
case 1 in the previous section with $t_0$ replaced by $v_0$. On the other
hand, the
expression
\r{a10}, which has no counterpart in the previous section, does satisfy the
Schr\"odinger equation.
Notice that $\vb s|s\hb=\int
d^nq\phi_{(b)}^*(q,t_0)\phi_{(b)}(q,t_0)$ for $\dot{\chi}(t)>0$. The
representation
with real time and the Schr\"odinger equation as boundary condition is
particular. One
may notice that translations in $v$ is performed by
 the unitary gauge operator
\be
&&U(b)\equiv e^{i[Q, \rho]}=e^{-ib\Pi_V}, \quad \rho=-b\baca.
\e{a11}
We have
\be
&&\Lambda(v-b)|\phi'\hb=0, \quad |\phi'\hb=U(b)|\phi\hb,
\e{a12}
which implies
\be
&\vb  q^i, t, v,  \ca,\baca|\phi'\hb&=\del(\Lambda(v-b))\varphi'(q, t,
v)=\nn\\&&=\del(v-b-v_0){1\over|\Lambda'(0)|}\varphi'(q, t,
v_0+b),
\e{a121}
and
\be
&&\varphi'(q, t,
v_0+b)=\varphi(q, t,
v_0).
\e{a122}
Thus, no Schr\"odinger equation is derived here in case a.

  We may also derive projected wave functions from the general
rules in section 4. Assuming $\dot{\chi}>0$  we find in case a for the partial
projection B (without imposing
$\Lambda=0$)
\be
&&\bar{\phi}(q,v,v_0)\equiv \int du d\baca d\ca \,
\Psi(q, iu, v, \ca, -i\baca)={1\over
|\Lambda'(v_0)|}\varphi(q,if(v-v_0),v_0).
\e{a13}
 (To set $\pet, \bapet$ to zero and $\pi=-H$ at $t=0$ is
equivalent to an integration over $\ca, \baca$ and $t=iu$.) The function
$f(v)=-i\chi^{-1}(-iv)$ is real and satisfies $f(0)=0$. Notice that
$\bar{\phi}(q,v,v_0)$ satisfies a Schr\"odinger equation with respect to $v$.
In case b we find instead the partially projected wave function
\be
&&\bar{\phi}(q,u,t_0)\equiv \int dt d\baca d\ca \,
\Psi(q, t, iu, \ca, -i\baca)={1\over
|\Lambda'(0)|}\varphi(q,t_0+f(u),0),
\e{a14}
which satisfies a
Schr\"odinger equation with respect to $u$.
The function $f(u)$ is here given by $f(u)=\chi^{-1}(u)-t_0$ and is  real
satisfying
$f(0)=0$.

Effectively the physical projected wave functions \r{a13} and \r{a14}
satisfy the
Schr\"odinger equation for imaginary and real times respectively which are
exactly
the reality properties of time chosen in the two representations!
If we also impose the gauge fixing conditions $\Lambda=0$
we find the physical
wave functions \r{a9} and \r{a10}.
 As in the previous section only projected wave functions with real
time represent the BRST singlets in the projection B. In case b we have
obviously
($\dot{\chi}>0$)
\be
&&\vb s|s\hb=\int d^n q \bar{\phi}^*(q,u,t_0)\bar{\phi}(q,u,t_0),
\e{a15}
from \r{a7}. However, there is no such
relation in case a for $\bar{\phi}(q,v,v_0)$ in
\r{a13}.

The physical wave functions
\r{a9} and
\r{a10} are also here directly related to
the wave function representations of the
$|\phi\hb$-states. We have in case a
\be
&&\vb q^i,  iu, v, \ca,
-i\baca|\phi\hb=\del(v-v_0){1\over|\Lambda'(v_0)|}\varphi(q,iu,v_0),
\e{a151}
 and in case b
\be
&&\vb q^i,  t, iu, \ca,
-i\baca|\phi\hb=\del(u){1\over|\Lambda'(0)|}\varphi(q,t,0).
\e{a152}
The projection prescription for these
wave functions given in subsection 4.2 requires
us to impose the dual boundary condition to the gauge fixing condition
$\Lambda=0$ which is
$\pi_v=0$ in case a. This is equivalent to
 an integration over $v$ which yields
${1\over|\Lambda'(v_0)|}\varphi(q,iu,v_0)$ which is equal to \r{a9} for
$u=0$, and which also only represents singlets at
$u=0$. In case b we should impose
$\pi_u=0$ which is equivalent to  an integration over $u$. We get then
${1\over|\Lambda'(0)|}\varphi(q,t_0,0)$,\ie exactly \r{a10}.

Consider the following gauge transformed singlet \r{a1}
\be
&&|\tilde{s}\hb=U(\beta)|s\hb=e^{{\ga\over2}[Q,\psi]}|\phi\hb,\nn\\
&&U(\beta)=e^{i\beta[Q,\rho]},\quad\rho\equiv\baca (V-v_0),
\e{a16}
where $\ga$ is a positive constant which depends on the real parameter
$\beta$.  We
have then
\be
&&\vb s|s\hb=\vb \tilde{s}|\tilde{s}\hb=\vb
\tilde{s}|U(b)|\tilde{s}\hb=\vb\phi|e^{\ga[Q,\psi]}|\phi'\hb=\nn\\
&&=\int d^{n+4}R d^{n+4}R'\vb\phi|R^*\hb\vb
R| e^{\ga[Q, \psi]} |R^{\prime *}\hb\vb R^{\prime}|\phi'\hb,
\e{a17}
where $R\equiv\{q^i, t, v, \ca, \baca \}$. From \r{a2} and \r{a12} we have
\be
&&\vb R^*|\phi\hb=\del(\Lambda(v^*))\varphi(q, t^*, v^*),\quad
\vb R^{\prime}|\phi'\hb=\del(\Lambda(v'-b))\varphi'(q', t', v'),
\e{a18}
where in distinction to \r{406} $\varphi$ depends on both $t$ and $v$. In
case a we
get therefore ($t=iu$)
\be
&&\vb\phi|e^{\ga[Q,\psi]}|\phi'\hb=\int d^nq d^nq' du du'\phi^*(q,
-iu, v_0)A(q, q';iu, iu'; v_0, v_0+b)\phi'({q'}, iu',  v_0+b),\nn\\
\e{a19}
where
\be
&&A(q, q';iu, iu'; v, v')\equiv
\int d\baca d\ca  d\baca' d\ca'  \vb q^i, iu, v,
\ca,
-i\baca|e^{\ga[Q, \psi]} | {q}^{\prime i}, iu', v', \ca',
-i\baca'  \hb=\nn\\&&=\ga
\dot{\chi}(iu)\del(u-u')\del(v-v'-i\ga\chi(iu))\del^n(q-q').
\e{a20}
This inserted into \r{a19} yields \r{a6}.

In case b $t$ is real,$v$
imaginary $(=iu)$ as well as $b$ in \r{a9} ($(=id)$). We get
\be
&&\vb\phi|e^{\ga[Q,\psi]}|\phi'\hb=\int d^nq d^nq' dt dt'\phi^*(q,
t, 0)A(q, q'; t,  t';0,  id)\phi'({q'}, t',  id),
\e{a21}
where
\be
&&A(q, q'; t,  t';iu, iu')
\equiv \int d\baca d\ca   d\ca' d\baca' \vb q^i, t, iu,
\ca,
-i\baca|e^{\ga[Q, \psi]} | {q}^{\prime i}, t', iu', \ca',
-i\baca'
\hb=\nn\\&&=\ga\dot{\chi}(t)\del(t-t')\del(u-u'-\ga\chi(t))\del^n(q-q').
\e{a22}
This inserted into \r{a21} yields \r{a7}.

The physical propagators should according to the rule in subsection 4.2 be
given by
\be
&&K(q,q';v,v')\equiv\int du du' A(q, q';iu, iu'; v,
v')=\sign(\ga\dot{\chi})\del^n(q-q')
\e{a23}
in case a and
\be
&&K(q,q';iu,iu')\equiv\int dt dt' A(q, q'; t,  t';iu,
iu')=\sign(\ga\dot{\chi})\del^n(q-q')
\e{a24}
in case b. Thus, although the above results are perfectly consistent,
\r{a23} and
\r{a24} are not the correct propagators for the Schr\"odinger equation.
Obviously
$v, v'$ in \r{a23} and $u, u'$ in \r{a24} are not time parameters. $K$ does not
depend   on them. In fact, \r{a23} and \r{a24} may be interpreted as the
physical
propagators in the limit of equal times. (Notice the delta-functions in
\r{a20} and
\r{a22} and that $\phi'({q'}, iu',  v_0+b)=\phi({q'}, iu',  v_0)$ in \r{a19} and
$\phi'({q'}, t',  id)=\phi({q'}, t',  0)$ in \r{a21} due to \r{a122}.)   To
get the
correct propagators for different times  we must use another gauge fixing
fermion. In fact, to make sure  always to obtain appropriate propagators we
must use
a gauge fixing fermion which is valid for any choices of
$|\phi\hb$-states. This is also what is required in the BFV path integral
formulation. This will be demonstrated for the above case within the path
integral
formulation below.

\setcounter{equation}{0}
\section{Path integrals for propagators}
In ordinary quantum mechanics we have the path integral representation
\be
&&\vb q'',t''|q', t'\hb=\vb q'',t'|U(t'',t')|q',
t'\hb=\nn\\&&=\int_{Path}\prod_t\frac{d^nq
d^np}{(2\pi)^n}\exp{\Big\{i\int^{t''}_{t'}dt\Big(p\cdot\dot{q}-H(p,q,t)\Big)
\Big\}}
\e{501}
where $U(t'',t')$ is given by \r{3104} with the Schr\"odinger Hamiltonian
$H_S(t)$
replaced by the operator
$H(P, Q, t)$.  $H(p, q, t)$ is  the Weyl symbol of
$H(P, Q, t)$ defined by
\be
&&H(p, q, t)=\frac{1}{(2\pi)^n}\int d^nu d^nv H(u,v, t)e^{-i(q\cdot u+p\cdot
v)}\nn\\
&&H(P, Q, t)=\frac{1}{(2\pi)^n}\int d^nu d^nv H(u,v, t)e^{-i(Q\cdot u+P\cdot
v)}.
\e{502}
The equality \r{501} may be obtained by means of the time slice formula. It is
strictly valid for bosonic coordinates in which case \r{501} is equivalent to
\r{409}. (For each fermionic coordinate
one has to remove a factor $2\pi$ in \r{501},\r{502} and insert an extra
factor $i$
in the exponentials (only in the first term in \r{501}).)

A similar procedure may
be used to obtain path integral representations for the inner products of
the BRST
singlets. Consider again the inner product  \r{403}.  According to \r{405}
we have
then
\be
&&\vb s|s\hb=\vb\phi|e^{\ga[Q,\psi]}|\phi'\hb=\int d^{n+4}R''
d^{n+4}R'\vb\phi'|R^{\prime\prime *}\hb\vb R^{\prime\prime}| e^{\ga[Q,
\psi]} |R^{\prime *}\hb\vb R^{\prime}|\phi\hb,
\e{504}
where $R'\equiv\{t', v', \pet', \bapet', q^{\prime i}\}$ etc. If we set
$\ga\equiv
\tau''-\tau'$ then we  obtain in  analogy to \r{501} the formal path integral
representation
\be
&&\vb R^{\prime\prime }| e^{\ga[Q, \psi]} |R^{\prime *}\hb\equiv\vb
R^{\prime\prime
}| e^{(\tau''-\tau')[Q,
\psi]} |R^{\prime *}\hb=\nn\\&&=\int_{Path}\prod_{\tau}\frac{d^{n+4} R
d^{n+4}
P}{(2\pi)^{n+2}}\exp{\Big\{i\int^{\tau''}_{\tau'}
d\tau\Big(P\cdot\dot{R}-i[Q,\psi]_W\Big)\Big\}},
\e{505}
where $R(\tau')\equiv R'$ and
$R(\tau'')\equiv R''$, and where $[Q,\psi]_W$ is the Weyl symbol of $[Q, \psi]$
defined in
\r{502} and in appendix B for fermions.
Notice that the $\tau$-parameter is introduced in a completely ad hoc manner.
Since $[Q, \psi]$
has no explicit
$\tau$-dependence it may be viewed as a conserved Hamiltonian with respect
to $\tau$.
Furthermore,  since the left-hand side is independent of
$\ga$ the right-hand side is independent of $\tau$. {\em Thus, reparametrization
invariance is implied.} (It is the unitary gauge transformation \r{401} which is
behind this.) As a general statement of expressions like \r{505} the
following may
be stated: The reality properties of the effective action depends on the chosen
representation. Typically it may be chosen either to be real or imaginary
\cite{Path}. The expression \r{505} is what is prescribed by the BFV
formalism if the
effective action is real and if $[Q,\psi]_W$ is the Poisson bracket of the BRST
charge and the gauge fixing factor. This is true in an appropriate
representation if
$Q$ and $\psi$ are sufficiently simple. This is the case here (see
subsection 6.2
below).

Let us calculate \r{505} explicitly. We consider then the representation in
which
$t$ is real and $v$ and $\pi_v$ are imaginary ($v=iu$, $\pi_{v}=-i\pi_u$
where $u,
\pi_u$ are real).   Furthermore, we choose
$\ca, \pet$ real and
$\baca, \bapet$ imaginary ($=-i\baca, i\bapet$; $\baca, \bapet$ real ). For
the gauge
fixing fermion
\r{31} we find
\be
&&\vb R^{\prime\prime }| e^{(\tau''-\tau')[Q, \psi]}
|R^{\prime *}\hb=\int_{Path}\prod_{\tau}\frac{d^{n}q  dt du d\baca d\ca
d^{n}p
d\pi
d\pi_u
d\pet
d\bapet}{(2\pi)^{n+2}}\times\nn\\&&
\exp{\Big\{i\int^{\tau''}_{\tau'}d\tau\Big(p_i\dot{q}^i+\pi\dot{t}+\pi_u
\dot{u}+i\ca\dot{\pet}+i\baca\dot{\bapet}-\la(\pi
+H(t))+i\la'\bapet\pet\Big)\Big\}},
\e{506}
where $\la(u)\equiv i\Lambda(iu)$ and $\la'(u)=-\Lambda'(iu)$. Notice that the
effective Lagrangian is real in the chosen representation. (Formally we may
choose
rather arbitrary complex eigenvalues of the ghosts. However, strictly this
is not
true when we use the time slice formulation since there are not the same
number of
slices in $\ca, \baca$ as in $\pet, \bapet$.) Integrations over the ghosts
$\ca$,
$\baca$, and
$\pi$,
$\pi_u$, yields
\be
&&\vb R^{\prime\prime}| e^{(\tau''-\tau')[Q, \psi]}
|R^{\prime *}\hb=\int_{Path}\prod_{\tau}\frac{d^{n}q  dt du
d^{n}p d\pet d\bapet
}{(2\pi)^{n}}\del(\dot{u})\del(\la-\dot{t})
\del(\dot{\bapet})\del(\dot{\pet})\times\nn\\&&
\exp{\Big\{i\int^{\tau''}_{\tau'}d\tau\Big(p_i\dot{q}^i
-\la H(t)+i\la'\bapet\pet\Big)\Big\}}.
\e{507}
The delta functions imply that $\pet$, $\bapet$, and $u$ and therefore also
$\la$ are
constants, and that
$t=\la\tau+c$ where
$c$ is another constant. Following the procedure of subsection 4.3 we define the
physical propagator by
\be
&&K(q'',q';t'',t')\equiv\int du' du'' d\pet' d\bapet'd\pet'' d\bapet''\vb
R^{\prime\prime}| e^{(\tau''-\tau')[Q, \psi]} |R^{\prime *}\hb.
\e{5071}
Inserting \r{507} and performing the integration over $u,\pet$ and $\bapet$
yield then
\be
&&K(q'',q';t'',t')=\varepsilon\vb q'',t''|q',
t'\hb,\quad\varepsilon=\sign((\tau''-\tau')\la'),
\e{508}
where the right-hand side is given by \r{501}. (Since we are integrating over
one more $u$ than $t$ in the time-slice procedure, also one delta function from
$\del(\la-\dot{t})$ contributes in the $u$-integrations which together with the
$\pet, \bapet$-integrations yields the sign factor.) Notice that
$t''-t'=(\tau''-\tau')\la$.  The result
\r{508} is exactly the same as
\r{409} which is expected since the treatment here is equivalent
to the one in subsection 4.3. Notice that the propagator contains all
information
including whether or not we have positive normed states.

\subsection{The general propagator}

The gauge fixing fermion \r{31} is relevant for $|\phi\hb$-states satisfying the
conditions \r{207} and \r{218}. It is more interesting to choose a
$|\phi\hb$-independent gauge fixing like \r{2152}, \ie
\be
&&\psi=\pet\Lambda(V)+\baca\chi(T).
\e{509}
The singlet $|s\hb=e^{[Q,\psi]}|\phi\hb$ is the same as \r{31} provided
$|\phi\hb$
satisfies
\r{207} and \r{218}, and provided $\Lambda(V)$ and $\chi(T)$ are linear.
However, this is not the case for the propagators. Choosing the same
representation as above  we find  the following propagator for the
gauge fixing fermion \r{509}
\be
&&\vb R^{\prime\prime }| e^{(\tau''-\tau')[Q, \psi]}
|R^{\prime *}\hb=\int_{Path}\prod_{\tau}\frac{d^{n}q  dt du d\baca d\ca
d^{n}p
d\pi
d\pi_u
d\pet
d\bapet}{(2\pi)^{n+2}} \exp{\{i\int_{\tau'}^{\tau''}d\tau
L_{eff}(\tau)\}},\nn\\&& L_{eff}(\tau)\equiv p_i\dot{q}^i+\pi\dot{t}+\pi_u
\dot{u}+i\pet\dot{\ca}+i\bapet\dot{\baca}-\la(\pi
+H(t))+i\la'\bapet\pet-\pi_u\chi(t)-i\dot{\chi}(t)\baca\ca,\nn\\
\e{510}
where as before $R'\equiv\{t', iu', \pet', i\bapet', q^{\prime i}\}$ etc. and
$\la(u)\equiv i\Lambda(iu)$. Defining the physical propagator as in
\r{5071} we find after integrations over the ghosts
$\pet$,
$\bapet$, and
$\pi$,
$\pi_u$
\be
&&K(q'',q';t'',t')=\int du' du''\int_{Path}\prod_{\tau}\frac{d^{n}q  dt du
d^{n}p d\baca d\ca \rho(\la')
}{(2\pi)^{n}}\del(\dot{u}-\chi(t))\del(\la(u)-\dot{t})\times\nn\\&&
\exp{\Big\{i\int^{\tau''}_{\tau'}d\tau\Big(p_i\dot{q}^i
+i{1\over\la'}\dot{\ca}\dot{\baca}-\la H(t)-i\dot{\chi}(t)\baca\ca\Big)\Big\}},
\e{511}
where $\rho(\la')=\la'$ is a measure factor. The conditions
\be
&&\dot{u}=\chi(t), \quad \dot{t}=\la(u)
\e{5111}
have well defined solutions if $\dot{\chi}(t)$ and $\la'(u)$ have definite signs
which we require. Let us for
simplicity  consider  the linear choice
\be
&&\chi(t)=\al(t-t_0), \quad \la(u)=\beta u,
\e{512}
where $\al$ and $\beta$ are real constants. The equations \r{5111} have then the
solutions
\be
&&u(\tau)=Ae^{c\tau}+Be^{-c\tau},\quad c\equiv\sqrt{\al\beta},\nn\\
&&t(\tau)=t_0+{c\over\al}\Big(Ae^{c\tau}-Be^{-c\tau}  \Big),
\e{513}
where $A$ and $B$ are constants. We have real solutions for arbitrary real
$A$ and
$B$ if $\al\beta>0$ and for real $A=B$ or imaginary $A=-B$ if $\al\beta<0$.
For the
linear choice \r{512} it is straight-forward to perform the integrations in
$u$ and
$\ca$,
$\baca$ in \r{511}. We find then exactly the solution \r{508}. This result
is what we
should expect since the gauge fixing fermion \r{509} yields exactly the same singlets
as
$\psi=\pet\Lambda(v)$ for linear $\chi(t)$ and $\Lambda(v)$ as was
mentioned above.
Thus, the physical wave functions are the same which means that the physical
propagators should be the same. However, still the result is somewhat
unexpected since
the $|\phi\hb$-states themselves do not enter the definition of the physical
propagator. Only the conditions on $|\phi\hb$ enter in the form of boundary
conditions. (In fact, the solution \r{508} is also valid for nonlinear
$\Lambda$ and
$\chi$.)

Now the gauge fixing fermion \r{509} is good for any  $|\phi\hb$-state. In
particular it is good for the $|\phi\hb$-state considered in section 5. A
corresponding physical propagator must be consistent with the conditions
\r{a2} on
$|\phi\hb$. This means that we should define the physical propagator as follows
\be
&&\left.K(q'',q';u'',u')=\int dt' dt''\vb R^{\prime\prime}| e^{(\tau''-\tau')[Q,
\psi]} |R^{\prime *}\hb\right|_{\pet'=\pet''=\bapet'=\bapet''=0}.
\e{514}
Integrations over $\ca, \baca$ and $\pi, \pi_u$ yields
($\rho(\dot{\chi})=\dot{\chi}$
is a measure factor)
\be
&&K(q'',q';u'',u')=\int dt' dt''\int_{Path}\prod_{\tau}\frac{d^{n}q  dt du
d^{n}p d\pet d\bapet \rho(\la')
}{(2\pi)^{n}}\del(\dot{u}-\chi(t))\del(\la(u)-\dot{t})\times\nn\\&&
\exp{\Big\{i\int^{\tau''}_{\tau'}d\tau\Big(p_i\dot{q}^i
-i{1\over\la'}\dot{\pet}\dot{\bapet}-\la H(t)+i{\la}'\bapet\pet\Big)\Big\}}.
\e{515}
and a further integration over $t$ and $\pet, \bapet$ yields then
\be
&&K(q'',q';u'',u')=\eta\vb q'',u''|q',
u'\hb,\quad\eta=\sign((\tau''-\tau')\dot{\chi}),
\e{516}
where $\vb q'',u''|q',
u'\hb$ with $u$ replaced by $t$ is given by \r{501} which is the
appropriate physical
propagator.

\subsection{Remarks on interpretations}
In all propagators above there appear delta functions implying the gauge fixing
\be
&&{dt\over d\tau}=\la(u).
\e{517}
Since the sign of $\tau''-\tau'$ is fixed by construction we have
$\sign\la=\sign(t''-t')$ for $\tau''>\tau'$ in agreement with previous results.
However, if we for some reason restrict the sign of $\la$ then
$\sign(\tau''-\tau')\propto\sign(t''-t')$ and
\be
&&K(q'',q';t'',t')\propto\sign(t''-t')\vb q'', t''|q', t'\hb,
\e{518}
which implies that $K$ no longer satisfies the Schr\"odinger equation. On
the other
hand such a restriction also implies that we consider both positive and negative
metric states for the BRST singlets. In order to restrict ourselves to positive
metric states we should choose positive sign of $\ve$ which here means that
$K$ only
is valid for $t''>t'$ or $t''<t'$ in which case the Schr\"odinger equation is
satisfied.
\subsection{Comparison with the standard BFV formulation}
The standard BFV path integral formulation of the propagator is given by the
right-hand side of \r{505} with $[Q, \psi]_W$ replaced by the Poisson
bracket $\{Q,
\psi\}$. In this formulation we obtain  exactly the right-hand side of
\r{510} for
the choice $\psi=\pet\la(u)+\baca\chi(t)$. Note that in the standard BFV
prescription
one should use real classical variables throughout in which case one always
obtain a
real effective Lagrangian. Let us summarize the differences between the two
approaches:\\ 1) $\la(u)$ is different from the classical function
$\Lambda(v)$ used
in the operator approach. In the linear case we have \eg $\la(u)=-\Lambda(u)$.\\
2) In the BFV formulation we may choose $\la(u-u_0)$ which is unnatural but
possible
within the operator approach. (It violates  manifest reality of the norms.) This
choice yields the same result as above: Projected propagators do not depend on
$u_0$.\\
3) In the BFV formulation one often considers $\tau$-dependent gauge fixing
$\chi(t)$. For instance, $t=\tau$ is a natural gauge fixing in the
classical theory.
A $\tau$-dependent $\chi(t)$ is very unnatural in the operator approach
although it
is in principle possible. In fact, \r{517} suggests \eg that we should choose
$t=\la\tau$ for $\la$ constant. (A similar gauge fixing has been proposed
for the
relativistic particle \cite{TG}.)\\ 4) All the above statements concern the
particular representation in which
$v$ and
$\baca$ and their conjugate momenta are imaginary. In the operator approach
we may
also consider other representations with complex time. Such representations are
unnatural but possible within the standard BFV formulation although there is no
general prescription for them there. The  argument for an
imaginary time and Euclidian field theory as due to indefinite metric space,
\ie the argument used in the present operator approach, was  in the literature
 first given in
\cite{AFIO}. Here we have shown that only the representation with real time
represent
the BRST singlets and the original theory.

In the general case the operator approach may even lead to an effective
Lagrangian
which depends on
$\hbar$. Such terms might appear in the Weyl symbol $[Q,\psi]_W$ and cannot
be argued
for within the standard BFV formulation. This  seems therefore to
constitute a crucial difference between the two approaches which has to be
clarified.

\setcounter{equation}{0}
\section{Conclusions}
In this paper we have performed a rather detailed investigation of the BRST
quantization of a simple class of reparametrization invariant theories
corresponding
to quantum mechanical systems, which we called  cohomological quantum
mechanics. The procedure used was a very precise operator version  of the
BFV formulation on inner product
spaces which prescribes the BRST singlets to be of the
form
$|s\hb\equiv e^{[Q,\psi]}|\phi\hb$ where $\psi$ is an appropriate hermitian
gauge
fixing fermion and where
$|\phi\hb$ satisfies conditions whose possible form are precisely
prescribed. There
are gauge invariant conditions as well as ghost and gauge fixings. They
determine the
boundary conditions in the wave function representation of $|s\hb$ as well as in
propagators. By means of these boundary conditions our analysis has
led us to propose the following general projection prescriptions:
\ben
\item Physical wave functions satisfying the Dirac quantization are
obtained from the
wave function representations of
$|s\hb$ by imposing boundary conditions  corresponding to the conditions on
$|\phi\hb$
except for the gauge fixings.
\item Gauge fixed physical wave functions, which not necessarily satisfy
the Dirac
quantization conditions,  are obtained from the above projected wave
function when the
boundary conditions corresponding to the gauge fixings are imposed. They
are also
obtained from the wave function representations of
$|\phi\hb$ by imposing the dual or conjugated boundary conditions to the gauge
fixings.
\item Physical propagators satisfying the Dirac quantization are obtained
from the
matrix representation of
$e^{[Q,\psi]}$ for an appropriate gauge fixing fermion $\psi$,  a
representation which
in general is equivalent to the conventional BFV path integral. One has
then to impose
the boundary conditions corresponding to the conditions on $|\phi\hb$
except for the
gauge fixings.
\een
For cohomological quantum mechanics we verified these rules explicitly. The
physical
wave functions in 1 and propagators in 3 were shown to   satisfy
the original Schr\"odinger equation exactly what we should have.  Property
3  was
shown to be true for gauge fixing fermions appropriate for a particular set of
conditions on
$|\phi\hb$, and for those that are independent of the conditions on
$|\phi\hb$. Their
path integral expressions were shown to agree with the projections obtained
from the
standard BFV prescription. The corresponding path integrals represent therefore
solutions to the Dirac quantization.  In \cite{HH} this was proposed to be
a general
property for a particular set of boundary conditions. (This was generalized to
reducible gauge theories in \cite{FHP}.) Our results suggest that this is a
general
property for all allowed boundary conditions. The understanding of these
boundary
conditions is also made more precise. (Reducible theories are considered in
\cite{Solv}.)

Our projected
wave functions may also depend on complex time since the   reality property
of time is
just a choice of representation. However, we have found that only wave
functions for
real time represent the BRST singlets. Although we expect that our
projection rules
are valid in general we need to verify this for many more models. In the
case of the
ordinary relativistic particle the rule 1 yields wave functions that satisfy the
Klein-Gordon equation. This may easily be extracted from
\cite{Proper}. That the corresponding propagators satisfy 3 is verified in
appendix C.

We hope that our treatment  has further elucidated how physical time
appears in a BRST
quantization of a reparametrization invariant theory.  Of course in gravity
and other more complicated theories everything is much more complex since
we then
 also have different topological sectors to deal with.\\ \\
\begin{appendix}
\section{Classical gauge invariant extensions}
Consider a classical gauge theory in which $\psi_s$, $s=1,\ldots,m$, are
the gauge
generators  satisfying a Lie algebra. ($\psi_s=0$ are first class
constraints.) To
every dynamical variable, $A$, there exists a gauge invariant extension in
the gauge,
$\chi^r=0$,
$r=1,\ldots,m$, given by (formula (3.13) in \cite{Pol})
\be
&&A_{(\chi)}=\int d\Om\: |\det{\{\chi^r(\Om),
\psi_s(\Om)\}}|\del^m(\chi(\Om))A(\Om),
\e{b1}
where $A(\Om), \chi^r(\Om)$ are the gauge transformed variable and gauge fixing
conditions, which explicitly depend on the group coordinates. The
integration is over
the group volume.

In  the cohomological models in the text there is only one gauge generator,
$\pi+H$,
and only one group coordinate. A gauge transformed dynamical variable $A$ is
here given by
\be
&&A(u)\equiv e^{-u Ad (\pi+ H(t))}A=e^{u(\dif_t-ad H-(\dif_tH)\dif_\pi)}A,
\e{b2}
where
\be
&&Ad H\equiv \{H,\cdot\}=ad H +{\dif H\over\dif t}{\dif\over\dif\pi},\nn\\
&&ad H\equiv \{H,\cdot\}_{red}={\dif H\over\dif q^i}{\dif \over\dif p_i}-{\dif
H\over\dif p_i}{\dif \over\dif q^i},
\e{b3}
where in turn $\{\;,\;\}$ is the Poisson bracket and $\{\;,\;\}_{red}$  the
reduced
Poisson bracket in which $t$ and $\pi$ are not dynamical. (It is {\em not}
the Dirac
bracket.)

 Let now
$\chi(t)=0$ be gauge fixing conditions that fix
$t$ to be
$t_0$ as in the text. From \r{b1} we have then that the gauge invariant
extension
$A_{(\chi)}$ of any dynamical variable $A$ is given by ($\pi+H(t)$ is
invariant under
group transformations here)
\be
&&A_{(\chi)}\equiv \int du\: |\det{\{\chi(t+u),
\pi+H(t)\}}|\del(\chi(t+u))A(u)=\nn\\
&&\left.=e^{-u(Ad \pi+Ad H)}A\right|_{u=t_0-t}.
\e{b5}
It is easy to see that $A_{(\chi)}$ is gauge invariant;
\be
&&(Ad \pi+Ad H(t))A_{(\chi)}=\{\pi+H(t), A_{(\chi)}\}=0.
\e{b6}
(Take \eg the derivative of $A_{(\chi)}$   with respect
to $t_0$ using the formal expression \r{b5}. One get then $(Ad \pi+Ad
H)A_{(\chi)}$
plus the derivative of $A_{(\chi)}$ with respect to
$t_0$ again.)  Eq.\r{b6} may be interpreted as follows: If we view $H(t)$ as the
Hamiltonian then  we have
\be
&&{dA_{(\chi)}\over dt}={\dif A_{(\chi)}\over \dif t} +\{A_{(\chi)},
H\}=-\{\pi+H,
A_{(\chi)}\}=0,
\e{b61}
which means that the gauge invariant extensions may be viewed as the
constants of
motion with respect to $t$.

We want now to derive the equations of motion with respect to $t_0$. We
need then the
following gauge invariant extensions
\be
&&\bar{q}^i(t_0)\equiv q^i_{(\chi)}\equiv \int du\: |\det{\{\chi(t+u),
\pi+H(t)\}}|\del(\chi(t+u))q^i(u)=\nn\\&&=\left.e^{u(\dif_t-ad
H(t))}q^i\right|_{u=t_0-t},\nn\\
&&\bar{p}_i(t_0)\equiv p_{i(\chi)}\equiv \int du\: |\det{\{\chi(t+u),
\pi+H(t)\}}|\del(\chi(t+u))p_i(u)=\nn\\&&=\left.e^{u(\dif_t-ad
H(t))}p_i\right|_{u=t_0-t},\nn\\
 &&\bar{H}(t_0)\equiv H_{(\chi)}\equiv \int du\: |\det{\{\chi(t+u),
\pi+H(t)\}}|\del(\chi(t+u))H(t,u)=\nn\\&&=\left.e^{u(\dif_t-ad
H(t))}H(t)\right|_{u=t_0-t},\nn\\
 &&t_{(\chi)}\equiv \int du\: |\det{\{\chi(t+u),
\pi+H(t)\}}|\del(\chi(t+u))(t+u)=t_0.
\e{b7}
The last relation states that the gauge invariant extension of time is
$t_0$, which
means that it is correct to look for the equations with respect to this
parameter.
Now the gauge invariant extensions are just the canonical transformations of the
original variables. This means that
\be
&&\bar{H}(t_0)=H(\bar{q}, \bar{p}, t_0).
\e{b8}
where $H({q},{p}, t)$ is the original Hamiltonian. Notice also that
$\left.\bar{q}^i(t_0)\right|_{t_0=t}=q^i$,
$\left.\bar{p}_i(t_0)\right|_{t_0=t}=p_i$, and
$\left.\bar{H}(t_0)\right|_{t_0=t}=H$. We have now  Hamilton's equations
\be
&&{d\bar{q}^i(t_0)\over dt_0}=\{\bar{q}^i(t_0),\bar{H}(t_0)\},\nn\\
&&{d\bar{p}_i(t_0)\over dt_0}=\{\bar{p}_i(t_0),\bar{H}(t_0)\}.
\e{b9}
Proof:
\be
&&{d\bar{q}^i(t_0)\over dt_0}=\left.e^{u(\dif_t-ad
H(t))}(\dif_t-ad
H(t))q^i\right|_{u=t_0-t}=\left.e^{u(\dif_t-ad
H(t))}\{q^i, H(t)\}\right|_{u=t_0-t}=\nn\\&&=\left.\{e^{u(\dif_t-ad
H(t))}q^i, e^{u(\dif_t-ad
H(t))}H(t)\}\right|_{u=t_0-t}=\{\bar{q}^i(t_0),\bar{H}(t_0)\}.
\e{b91}
In the special case  when $H$ has no explicit time dependence we have in
particular
\be
&&\bar{H}(t_0)=H(\bar{q}, \bar{p})=H(q, p), \quad  \bar{q}^i(t_0)=e^{-(t_0-t)ad
H}q^i, \nn\\&& \bar{p}_i(t_0)=e^{-(t_0-t)ad
H}p_i,
\e{b10}
from which \r{b9} follows trivially. Eq.\r{b9} are the correct classical
Hamilton's
equations since the Poisson bracket may be expressed in terms of the canonical
coordinates
$\bar{q}^i$, $\bar{p}_i$.

\setcounter{equation}{0}
\section{Properties of our ghost states}
Following the conventions of \cite{Fermi} we define ghost eigenstates as
follows (we
put hats on the operators here for clarity)
\be
&&|\ca, \baca\hb\equiv e^{-\ca\hpet-\baca\hbapet}|0\hb_{\ca\baca},\nn\\
&&|\pet, \bapet\hb\equiv e^{-\pet\hca-\bapet\hbaca}|0\hb_{\pet\bapet},
\e{c1}
where the vacuum states are normalized as follows
\be
&&_{\pet\bapet}\vb0|0\hb_{\ca, \baca}=\,_{\ca,
\baca}\vb0|0\hb_{\pet\bapet}=1,\nn\\
&&_{\pet\bapet}\vb0|\hca\hbaca|0\hb_{\pet\bapet}=\,_{\ca,
\baca}\vb0|\hpet\hbapet|0\hb_{\ca, \baca}=i.
\e{c2}
This implies ($\vb
\ca^*,\baca^*|\equiv(|\ca,
\baca\hb)\dagg$)
\be
&&\vb \ca^*,\baca^*|\ca',\baca'\hb=i
\del(\ca^*-\ca')\del(\baca^*-\baca'),\nn\\
&&\vb \ca^*,\baca^*|\pet, \bapet\hb=e^{-\pet\ca^*-\bapet\baca^*}.
\e{c3}
For $\ca$ real and $\baca$  imaginary ($=\pm i\baca$) we have in particular
\be
&&\vb
\ca,i\baca|\ca',i\baca'\hb=\del(\baca-\baca')\del(\ca-\ca'),\nn\\
&&\vb \ca,-i\baca|\ca',-i\baca'\hb=\del(\ca-\ca')\del(\baca-\baca'),
\e{c4}
where all coordinates are real. The corresponding completeness relations are
\be
&&\int |\ca, i\baca\hb d\ca d\baca\vb\ca, i\baca|=\bett,\nn\\
&&\int |\ca, -i\baca\hb d\baca d\ca \vb\ca, -i\baca|=\bett.
\e{c5}
Note that $\int d\ca \ca=i$ and $\int d\baca \baca=i$ in the conventions we are
using
\cite{Fermi}. (For $\int d\ca \ca=\int d\baca \baca=1$ one has to do the
replacement $d\ca d\baca\leftrightarrow d\baca d\ca$.)  The properties of
$|\pet, \pm i\bapet\hb$ are also given by
\r{c4} and \r{c5} with $\ca, \baca$ replaced by $\pet, \bapet$. Note that
\be
&&\vb \ca,\pm i\baca|\pet, \mp i\bapet\hb=e^{-\pet\ca-\bapet\baca},
\e{c6}
according to \r{c3}.

The Weyl transform $A_W$ (or Weyl symbol) of an even operator $A$ is defined by
\be
&&A(\hpet,\hbapet,\hca,\hbaca)=\int d\beta_1
d\beta_2 d\al_1 d\al_2 A_W(\al_1, \al_2,
\beta_1, \beta_2)\Delta(\al_1, \al_2,
\beta_1, \beta_2),\nn\\
&&\Delta(\al_1, \al_2,
\beta_1, \beta_2)\equiv \nn\\&&\int d\la_2 d\la_1 d\xi_2 d\xi_1
\exp{\{(\al_1-\hpet)\la_1+(\al_2-\hbapet)\la_2+
(\beta_1-\hca)\xi_1+(\beta_2-\hbaca)\xi_2\}}.
\e{c7}
From this definition we get (cf \cite{Path})
\be
&&\vb \pet, i\bapet|A(\hpet,\hbapet,\hca,\hbaca)|\pet',
 i\bapet'\hb=\nn\\&&\int d\baca
d\ca A_W(\half(\pet+\pet'), i\half(\bapet+\bapet'), \ca,
\baca)e^{-\ca(\pet-\pet')}e^{-\baca(\bapet-\bapet')},\nn\\
&&\vb \ca, -i\baca|A(\hpet,\hbapet,\hca,\hbaca)|\ca', -i\baca'\hb=\nn\\&&
\int d\pet
d\bapet A_W(\pet, \bapet, \half(\ca+\ca'),
-i\half(\baca+\baca'))e^{-\pet(\ca-\ca')}e^{-\bapet(\baca-\baca')}.
\e{c8}
These expressions determine the Weyl symbol $[Q,\psi]_W$ in the path
integrals in
section 6.\\ \\

\setcounter{equation}{0}
\section{Application to the relativistic particle}
 A free relativistic particle satisfies
the mass shell condition
\be
&&p^2+m^2=0.
\e{d1}
 Efter quantization this constraint yields the Klein-Gordon
equation. The appropriate BFV-BRST charge is then
\be
&&Q=\half(P^2+m^2)\ca+\Pi_V\bapet,
\e{d2}
which has the same form as for cohomological quantum mechanics in the text
except that
$\Pi+H$ is replaced by $\half(P^2+m^2)$. The canonical set of variables are
$(X^{\mu},
P^{\mu})$, $(V,\Pi_V)$, $(\ca, \pet)$ and $(\baca, \bapet)$ which satisfy the
commutator algebra \r{204} and \r{206} and
\be
&&[X^{\mu}, P^{\nu}]=i\eta^{\mu\nu},
\e{d3}
where $\eta^{\mu\nu}$ is the Minkowski metric with signature $(-,+,+,+)$. As in
cohomological quantum mechanics  it is natural to use the representation
in which $V$ and $\bapet$ have imaginary eigenvalues. In \cite{Proper} and
\cite{Gauge} two forms of the BRST singlets for the relativistic particle
were given.
Let $\psi_1(x, iu, \pet, i\bapet)$ be the wave function representation of
the singlet
in \cite{Proper} where $iu$ and $i\bapet$ are eigenvalues of the operators
$V$ and
$\bapet$. The appropriate boundary conditions are here
$\pi_u=\ca=\baca=0$. It is then straight-forward to check that
$\phi_1(x)\equiv\int du d\pet d\bapet
\psi_1(x, iu, \pet, i\bapet)$ satisfies the Klein-Gordon equation. Let
$\psi_2(x, iu,
\pet, i\bapet)$ be the wave function representation of the singlet in
\cite{Gauge}.
The appropriate boundary conditions are then $p^2+m^2=\pet=\bapet=0$.  The
projected
wave function
$\phi_2(x)\equiv\int dx^0 \psi_2(x, iu, \pet=0, i\bapet=0)$ satisfies then the
Klein-Gordon equation where time
$x^0$ may be identified with $a+u$ where the parameter $a$ enters the
chosen gauge
fixing fermion which is $\psi=\baca(X^0-a)$ here. To be precise we should
define the
physical wave function by
\be
&&\tilde{\phi}_2(x)\equiv\int
d^4p\del(p^2+m^2)e^{ip\cdot x}\tilde{\psi}_2(p, iu, \pet, i\bapet),
\e{d31}
where
$\tilde{\psi}_2$ is the Fourier transform of ${\psi}_2$ with respect to
$x^{\mu}$.
Also $\tilde{\phi}_2(x)$ satisfies the Klein-Gordon equation. ${\phi}_2(x)$ and
$\tilde{\phi}_2(x)$ are equivalent apart from a factor $p^0$ at $x^0=a+u$.

Let us now consider the general propagator
\be
&&\vb R''|e^{(\tau''-\tau')[Q, \psi]}|R^{\prime *}\hb,
\e{d4}
where $Q$ is given by \r{d2} and where the gauge fixing fermion $\psi$ is
\be
&&\psi=\al\pet(V-v_0)+\beta\baca(X^0-a),
\e{d5}
¨where $\al,\; \beta, \;v_0$ and $a$ are real parameters. We find
\be
&&[Q, \psi]=\half\al(P^2+m^2)(V-v_0)-i\al\bapet
\pet+\beta\pi(X^0-a)-i\beta\ca\baca
P^0.
\e{d6}
If $R=\{x^{\mu}, iu, \ca, -i\baca\}$ in \r{d4} where $iu$ and $-i\baca$ are
eigenvalues of the operators $V$ and $\baca$ then \r{d4} has the path integral
representation \r{510} where the effective Lagrangian is (choosing $v_0=0$
and the
eigenvalues of $\Pi_V$ to be $-i\pi_u$)
\be
&&L_{eff}(\tau)\equiv
p_{\mu}\dot{x}^{\mu}+\pi_u\dot{u}+i\pet\dot{\ca}+i\bapet\dot{\baca}+
\half\al(p^2+m^2)u-\nn\\&&-i\al\bapet\pet-\beta\pi_u(x^0-a)+i\beta\ca\baca
p^0,
\e{d7}
which also may be obtained from the BFV path integral prescription.
Let us first calculate the projection to the physical propagator for the
boundary
conditions $\pi_u=\ca=\baca=0$. We obtain ($\rho(\al)=\al$ is a measure factor)
\be
&&\left.K(x'';x')\equiv\int du' du'' \vb
R''|e^{(\tau''-\tau')[Q, \psi]}|R^{\prime *}\hb\right|_{\baca'=\ca'=\baca''
=\ca'=0}=\nn\\&&=
\int du' du''\int_{Path}\prod_{\tau}\frac{d^{4}x   du
d^{4}p d\baca d\ca \rho(\al)
}{(2\pi)^{4}}\del(\dot{u}-\beta(x^0-a))\times\nn\\&&
\left.\exp{\Big\{i\int^{\tau''}_{\tau'}d\tau\Big(p_{\mu}\dot{x}^{\mu}
+i{1\over\al}\dot{\ca}\dot{\baca}+i\beta
p^0\ca\baca+\half\al(p^2+m^2)u\Big)\Big\}}\right|_{\baca'=\ca'=\baca''
=\ca'=0}=\ve
P(x'';x'),\nn\\&&\ve\equiv
\sign(\al(\tau''-\tau')),\quad P(x'';x')\equiv\int
d^4p\del(\half(p^2+m^2))
e^{ip_{\mu}(x^{\prime\prime\mu}-x^{\prime\mu})}.
\e{d8}

If we instead impose the boundary
conditons $p^2+m^2=\pet=\bapet=0$  we find
\be
&&K'(x'';x')\equiv\int d^4p'
d^4p''\del(\half({p''}^2+m^2))
\del(\half({p'}^2+m^2))e^{ip''\cdot x''-ip'\cdot
x'}\times\nn\\&&\left.\vb R''|e^{(\tau''-\tau')[Q,
\psi]}|R^{\prime *}\hb\right|_{\bapet'=\pet'=\bapet'' =\pet'=0}=
\int d^4p'
d^4p''\del(\half({p''}^2+m^2))
\del(\half({p'}^2+m^2))\times\nn\\&&e^{ip''\cdot
x''-ip'\cdot x'}
\int_{Path}\prod_{\tau}\frac{d^{4}x   du
d^{4}p d\pet d\bapet \rho(\beta p^0)
}{(2\pi)^{4}}\del(\dot{u}-\beta(x^0-a))\times\nn\\&&
\left.\exp{\Big\{i\int^{\tau''}_{\tau'}
d\tau\Big(-\dot{p}_{\mu}{x}^{\mu}
-i{1\over\beta p^0}\dot{\pet}\dot{\bapet}-i\al
\bapet\pet+\half\al(p^2+m^2)u\Big)
\Big\}}\right|_{\bapet'=\pet'=\bapet'' =\pet'=0}
=\nn\\&&=\ve'
P'(x'';x'),\quad\ve'\equiv\sign(\beta
(\tau''-\tau')),\nn\\&&P'(x'';x')\equiv\int
d^4p\,\sign(p^0)\del(\half(p^2+m^2))
e^{ip_{\mu}(x^{\prime\prime\mu}-x^{\prime\mu})}.
\e{d9}
Notice that the negative energy contribution
enters with wrong sign in $P'(x'';x')$
as compared to
$P(x'';x')$ in
\r{d8}. This means that the last projections cannot lead to positive normed
states
which is the case in \r{d8}. This is consistent with the results of
\cite{Proper} and
\cite{Gauge}. Notice that both $P(x'';x')$ in \r{d8} and
$P'(x'';x')$ in \r{d9}
satisfy the Klein-Gordon equation.

We may also use a different state
space representation in which $X^0$ and $P^0$ have
imaginary eigenvalues instead of
$V$ and $\Pi_V$. We have then the general propagator
\r{d5} with $R=\{{\bf x}, ix^0, v, \ca, -i\baca\}$ for which the path integral
representation \r{510} is valid with the effective Lagrangian ($a=0$ here)
\be
&&L_{E\; eff}(\tau)\equiv
p_{\mu}\dot{x}^{\mu}+\pi_v\dot{v}+
i\pet\dot{\ca}+i\bapet\dot{\baca}-
i\half\al(p^2+m^2)(v-v_0)-\nn\\&&-
i\al\bapet\pet+\beta\pi_v x^0-\beta\ca\baca
p^0,
\e{d10}
where `E' denotes Euclidean since the
metric for $x^{\mu}$ and $p^{\mu}$ is Euclidean.
This effective Lagrangian is not
entire real and it cannot be obtained from the
standard BFV path integral prescription.
 For the boundary
conditions
$\pi_v=\ca=\baca=0$ we get here the physical propagator
\be
&&\left.K_E(x'';x')\equiv\int dv' dv'' \vb
R''|e^{(\tau''-\tau')[Q, \psi]}|R^{\prime *}\hb\right|_{\baca'=\ca'=\baca''
=\ca'=0}=\nn\\&&=
\int dv' dv''\int_{Path}\prod_{\tau}\frac{d^{4}x   dv
d^{4}p d\baca d\ca \rho(\al)
}{(2\pi)^{4}}\del(\dot{v}+\beta(x^0-a))\times\nn\\&&
\left.\exp{\Big\{i\int^{\tau''}_{\tau'}d\tau\Big(p_{\mu}\dot{x}^{\mu}
-i{1\over\al}\dot{\ca}\dot{\baca}-\beta
p^0\ca\baca-i\half\al(p^2+m^2)(v-v_0)
\Big)\Big\}}\right|_{\baca'=\ca'=\baca''
=\ca'=0}=\nn\\&&=\ve
P_E(x'';x'),\quad\ve\equiv
\sign(\al(\tau''-\tau')),\quad P_E(x'';x')\equiv2\int
d^4p\frac{e^{ip_{\mu}
(x^{\prime\prime\mu}-x^{\prime\mu})}}{p^2+m^2}.
\e{d11}
This is only valid if the Lagrange
multiplier has the range $(v_0,\infty)$ for
$\al(\tau''-\tau')<0$, and $(-\infty, v_0)$ for $\al(\tau''-\tau')>0$. (We
cannot
have infinite range and these ranges are natural.) Such a restriction of
the Lagrange
multiplier has to be done at the very
beginning by choosing variables $\om, p_{\om}$
defined by either
$v=v_0+e^{\om}, p_v=e^{-\om}p_{\om}$ or
$v=v_0-e^{\om}, p_v=-e^{-\om}p_{\om}$ where $\om$ has infinite range. (Such
parametrization of the relativistic particle was made in \cite{NW}.) The
restriction
in the ranges for the Lagrange multiplier above seem a little ad hoc. One should
therefore start from a geometric argument for why $v-v_0$ should be positive or
negative
\cite{TG,JG}. Then one finds that the Euclidean representation above must
be used. In
\r{d11} the sign of
$\ve$ is not directly related to positive or negative metric states. We may have
either or depending on whether or not the wave functions of the BRST
singlets are
even or odd in $p^0$
\cite{Proper}.  The propagator $P_E(x'';x')$ is a Wick rotated Feynman
propagator
which does not satisfy the Klein-Gordon equation.  The Klein-Gordon
equation is only
satisfied if we restrict the sign of $x^{\prime\prime 0}-x^{\prime 0}$
which in a way
is what the interpretation in subsection 6.2 suggests us to do since the
range of the
Lagrange multiplier is restriced here. For positive or negative time differences
we have then propagators for only one branch of the solutions of the the
mass shell
condition. Notice that in this Euclidean case we cannot impose the boundary
conditions
$p^2+m^2=\pet=\bapet=0$  since the first equality then implies $p^{\mu}=m=0$.

Contrary to the case for cohomological quantum mechanics we have obtained three
different physical propagators here. Their difference seems to be  entirely
due to a
different treatment of the negative energy solution whose appearance is due to a
nontrivial topology.

\end{appendix}

\end{document}